\documentclass[fleqn,usenatbib]{mnras}

\usepackage{newtxtext,newtxmath}

\usepackage[T1]{fontenc}
\usepackage[normalem]{ulem}

\DeclareRobustCommand{\VAN}[3]{#2}
\let\VANthebibliography\thebibliography
\def\thebibliography{\DeclareRobustCommand{\VAN}[3]{##3}\VANthebibliography}

\usepackage{graphicx}	
\usepackage{amsmath}	

\newcommand{\lp}{\left(}
\newcommand{\rp}{\right)}
\newcommand{\dd}{{\rm d}}
\newcommand{\surf}{{\rm{surf}}}

\newcommand{\ns}{{\rm{NS}}}
\newcommand{\Msun}{\ensuremath{\rm{M}_\odot}}

\title[Baryon Ejection in Magnetar Flares]{Dynamics of baryon ejection in magnetar giant flares: implications for radio afterglows, \textit{r}-process nucleosynthesis, and fast radio bursts}

\author[Cehula, Thompson, \& Metzger]{
Jakub Cehula,$^{1}$\thanks{E-mail: jakub.cehula@mff.cuni.cz}
Todd A. Thompson,$^{2,3,4}$
and Brian D. Metzger$^{5,6}$
\\
$^{1}$Institute of Theoretical Physics, Faculty of Mathematics and Physics, Charles University, V Hole\v{s}ovi\v{c}k\'{a}ch 2, Prague, 18000, CZ\\
$^{2}$Department of Astronomy, Ohio State University, 140 West 18th Avenue, Columbus, OH 43210, USA\\
$^{3}$Center for Cosmology \& Astro-Particle Physics, Ohio State University, 191 West Woodruff Ave., Columbus, OH 43210, USA\\
$^{4}$Department of Physics, Ohio State University, 191 West Woodruff Ave., Columbus, OH 43210, USA\\
$^{5}$Department of Physics and Columbia Astrophysics Laboratory, Columbia University, New York, NY 10027, USA\\
$^{6}$Center for Computational Astrophysics, Flatiron Institute, 162 5th Ave., New York, NY 10010, USA
}

\date{Accepted XXX. Received YYY; in original form ZZZ}

\pubyear{2023}

\begin{document}
\label{firstpage}
\pagerange{\pageref{firstpage}--\pageref{lastpage}}
\maketitle

\begin{abstract}
We explore the impact of a magnetar giant flare (GF) on the neutron star (NS) crust, and the associated baryon mass ejection.  We consider that sudden magnetic energy dissipation creates a thin high-pressure shell above a portion of the NS surface, which drives a relativistic shockwave into the crust, heating a fraction of these layers sufficiently to become unbound along directions unconfined by the magnetic field.  We explore this process using spherically-symmetric relativistic hydrodynamical simulations.  For an initial shell pressure $P_{\rm GF}$ we find the total unbound ejecta mass roughly obeys the relation $M_{\rm{ej}}\sim4-9\times10^{24}\:\rm{g}\:(P_{\rm GF}/10^{30}\:\rm{ergs}\:\rm{cm}^{-3})^{1.43}$.  For $P_{\rm{GF}}\sim10^{30}-10^{31}\:\rm{ergs}\:\rm{cm}^{-3}$ corresponding to the dissipation of a magnetic field of strength $\sim10^{15.5}-10^{16}\:\rm{G}$, we find $M_{\rm{ej}}\sim10^{25}-10^{26}\:\rm{g}$ with asymptotic velocities $v_{\rm{ej}}/c\sim0.3-0.6$ compatible with the ejecta properties inferred from the afterglow of the December 2004 GF from SGR 1806-20.  Because the flare excavates crustal material to a depth characterized by an electron fraction $Y_e\approx0.40-0.46$, and is ejected with high entropy and rapid expansion timescale, the conditions are met for heavy element $r$-process nucleosynthesis  via the alpha-rich freeze-out mechanism.  Given an energetic GF rate of roughly once per century in the Milky Way, we find that magnetar GFs could be an appreciable heavy $r$-process source that tracks star formation.  We predict that GFs are accompanied by short $\sim$minutes long, luminous $\sim10^{39}\:\rm{ergs}\:\rm{s}^{-1}$ optical transients powered by $r$-process decay ({\it"nova brevis"}), akin to scaled-down kilonovae.  Our findings also have implications for the synchrotron nebulae surrounding some repeating fast radio burst sources.
\end{abstract}

\begin{keywords}
hydrodynamics -- nuclear reactions, nucleosynthesis, abundances -- shock waves -- stars: magnetars -- stars: winds, outflows -- fast radio bursts
\end{keywords}



\section{Introduction}

Magnetars are neutron stars (NSs) with  surface magnetic field strengths $\gtrsim 10^{14} \: \rm{G}$ \citep[e.g.,][]{duncan1992,usov1992,Kouveliotou98,kaspi2017} and relatively slow rotation periods of several seconds or longer.  Although the physical origin of their strong magnetic fields remains uncertain, Galactic magnetars are likely produced in at least tens of percent of core-collapse supernovae \citep[e.g.,][]{beniamini2019}.  

One of the hallmarks of magnetars is their transient outbursts, as manifested through a spectrum of hard X-ray and soft gamma-ray bursts \citep[e.g.,][]{gogue1999,gogue2000}.  Based on the total energy released, the bursts can be divided into three categories: short bursts (isotropic energies $E \lesssim 10^{41} \: \rm{ergs}$), intermediates flares ($E \sim 10^{41}-10^{43} \: \rm{ergs}$), and giant flares (GFs) ($E \sim 10^{44}-10^{47} \: \rm{ergs}$).  See, e.g., the reviews by \citet{turolla2015,kaspi2017}.  Thus far, only a handful of GFs have been observed in the Local Group \citep[e.g.,][]{mazets1979,hurley1999,hurley2005}, with some evidence for GFs outside of the Local Group \citep[e.g.,][]{Tanvir+05,ofek2006,svinkin2021,minaev2020,trigg2023}.

Perhaps the most notable example is the GF that occurred in December 2004 from SGR 1806-20 \citep[e.g.,][]{palmer2005,hurley2005}, an event which released $\sim 2-4 \times 10^{46} \: \rm{ergs}$ in hard X-rays and soft-gamma rays during the initial bright spike that lasted for $\approx 0.2-0.5 \: \rm{s}$, was followed by a decaying minutes-long tail modulated at the NS rotation period, and a $\sim$ hour-long hard X-ray afterglow that peaked $\sim 600 - 800$ s after the initial spike \citep[e.g.,][]{mereghetti2005,frederiks2007}.  The GF produced a radio afterglow observed with multiple facilities in the following weeks and months \citep[e.g.,][]{cameron2005}.  The light curve shows a steep decay ($\propto t^{-2.7}$) around $9$ days after the initial spike \citep{gaensler2005}, followed by a rebrightening around day $25$ lasting for about a week \citep{gelfand2005}.  Motion of the radio centroid and its polarization properties point to an asymmetric predominantly one-sided outflow \citep{taylor2005}.  A dynamical model capable of explaining the radio afterglow was proposed by \citet{gelfand2005,granot2006}, in which the early steep decline phase arises from baryon-rich ejecta colliding with a shell surrounding a pre-existing cavity, while the rebrightening occurs as the comoving ejecta and shell decelerate upon sweeping up more material in the ambient medium.  Most of the ejecta energy resides in material expanding at mildly relativistic speeds, with an initial velocity $\sim 0.7 c$, kinetic energy $\sim 10^{44.5}-10^{46} \: \rm{ergs}$, and total mass $\sim 10^{24.5}-10^{26} \: \rm{g}$, for an assumed distance of $15 \: \rm{kpc}$ \citep{granot2006}.  This scenario is also broadly consistent with the observed hard X-ray afterglow \citep{mereghetti2005}. Additional modeling of the optical and radio afterglows generated as magnetar flares interact with their environments was presented in \citep{Margalit+20,wei2023}, with \citet{wei2023} emphasizing the potential to detect extragalactic flares with current and future radio telescopes.

An alternative interpretation of the observations is that GFs are relativistic coronal mass ejections, in rough analogy with \emph{solar flares} \citep[e.g.,][]{lyutikov2006,lyutikov2015,mehta2021}.  In this picture, GFs are purely magnetospheric events that produce strongly relativistic, strongly magnetized, and baryon-poor ejecta.  \citet{lyutikov2006} constrains the mass ejected during the December 2004 GF to be $\lesssim 10^{22} \: \rm{g}$.  This model can reproduce the late-time behaviour of the radio light curve, but (at least thus far) provides no explanation for the radio rebrightening \citep{mehta2021}.  

In the baryon-ejection scenario of \citet{gelfand2005,granot2006}, the sudden energy release responsible for the GF results from the gradual build-up of magnetic stresses in the NS crust, released in a sudden \emph{starquake}/\emph{crustquake} \citep[e.g.,][]{thompson1995,thompson2001gf,perna2011,lander2015}. Note, however, that the dichotomy between the \emph{solar flare} and the \emph{starquake}/\emph{crustquake} paradigm is not clearly defined \citep[e.g.,][]{sharma2023}.  Recently, \citet{demidov2023}, building on earlier work by \citet{thompson1995}, considered that a portion of the energy released during the initial bright spike, is trapped in a radiatively cooling electron-positron pair fireball in the NS magnetosphere.  In such a strong magnetic field, the vacuum becomes birefringent and the photons are split into ordinary and extraordinary modes.  The ordinary mode interacts strongly with matter while the extraordinary mode interacts only weakly.  This allows for material to be ablated from the NS surface by irradiation, with \citet{demidov2023} finding that $\sim 10^{18} \: \rm{g}$ can be ejected this way over $\sim 100 \: \rm{s}$.  However, this mass-loss phase, which occurs during the minutes-long tail/fireball stage of the GF, is not sufficient to explain the mildly relativistic ejecta of mass $\gtrsim 10^{24.5} \: \rm{g}$ inferred by \citet{granot2006}, which instead is more likely to have been ejected during the initial stages of the flare when the magnetosphere is still open.  We propose a model for such a prompt ejecta phase in the present paper.  

The ejection of baryonic material during magnetar GFs would have consequences for a number of topics beyond just the phenomenology of Galactic magnetars.  In particular, NS crust material is neutron-rich, such that the decompression of GF ejecta into space may give rise to the conditions necessary for the creation of heavy elements via the rapid neutron-capture process ($r$-process).  The origin of the $r$-process is a long-standing mystery in nuclear astrophysics \citep{Burbidge+57,Cameron57,lattimer1974}.  Additionally, magnetars are considered the most likely central engines for fast radio bursts (FRBs; \citealt{lorimer2007,Lyubarsky14,Metzger+17,Kumar+17,Beloborodov17,Bochenek+20,CHIME+20}).  These models require large-quantities of baryon-rich transrelativistic ejecta to explain the high rotation measures inferred from the synchrotron nebulae that surround particularly active FRB sources \citep{margalit2018,Zhao&Wang21,Metzger&Sridhar22}, such as FRB 121102 (e.g., \citealt{Michilli+18}) and FRB 190520B (e.g., \citealt{Niu+22}).

\subsection{Physical Picture for Baryon Ejection in Magnetar Giant Flares}
\label{sec:overview}
All of the above motivates us to consider the hypothesis that the sudden energy release above the NS surface in a GF, drives a strong relativistic shock into the outer layers of the star that unbinds baryonic material from the NS crust.

\begin{figure*}
	\includegraphics[width=\textwidth]{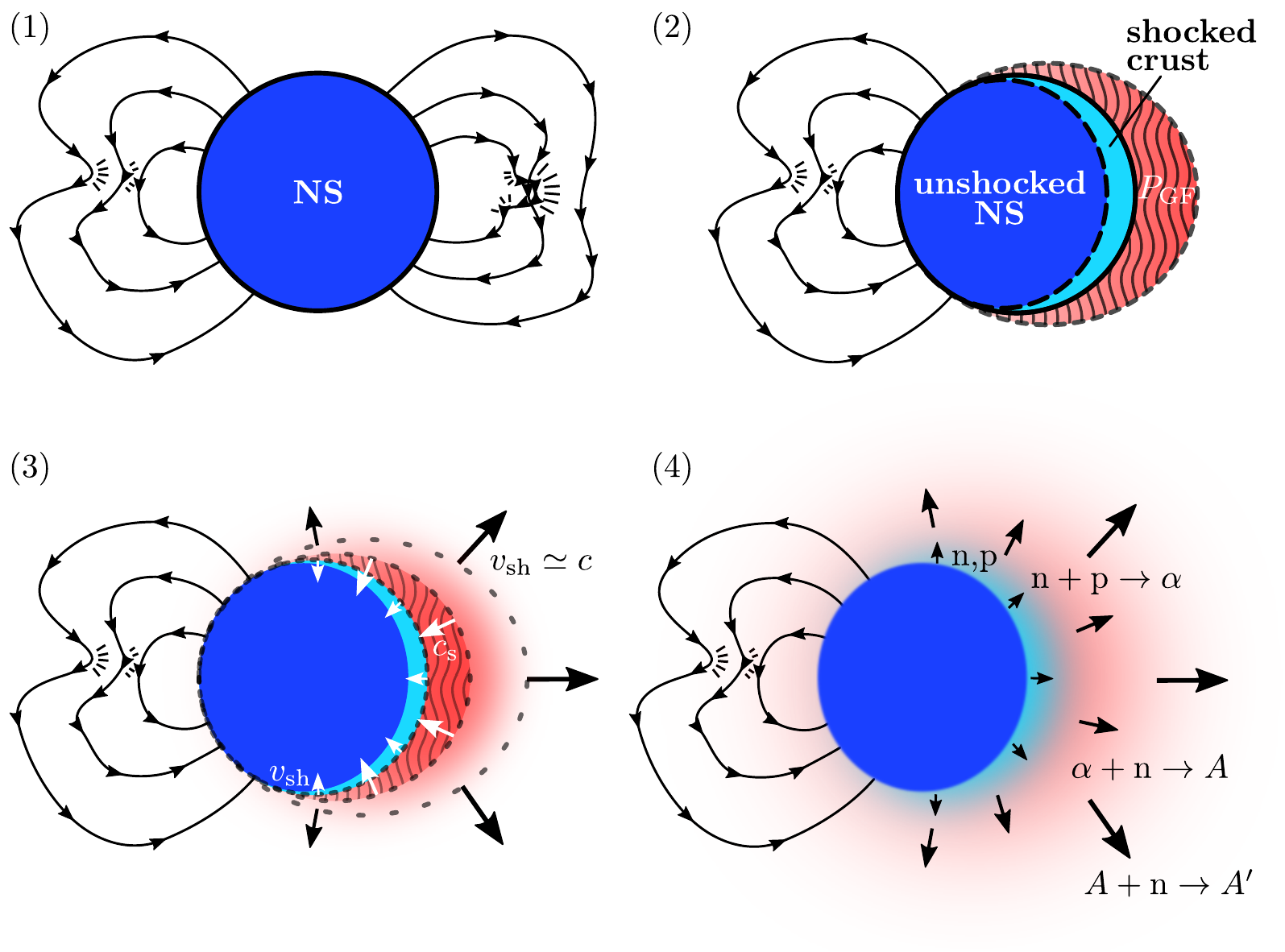}
    \caption{Schematic illustration of the envisioned process of baryonic mass ejection following a magnetar GF (see also Sec.~\ref{sec:analytic}).  (1) A sudden release of magnetic energy in the magnetosphere results in (2) the creation of a thin high-pressure layer ($P_{\rm GF}$; {\it red shell}) above a portion of the NS surface (right hand side in the diagram); (3) The high pressure drives two strong shocks, one into the surrounding low-density magnetosphere propagating at $v_{\rm sh} \simeq c$ and a second into the crust of the NS ({\it dark blue}).  Eventually, the inward-propagating shock stalls (typically meters to hundreds of meters below the original NS surface), and the pressure drop associated with the outward-expanding shock wave is communicated at the sound speed $c_{\rm s}$ to the inner edge of the high-pressure shell; (4) The shocked neutron-rich crustal layers ({\it teal shell}), photodissociated into free nucleons by the shock, re-expands into space.  A fraction of this material now possesses enough energy to reach the escape speed and become unbound.  These layers possess sufficiently high entropy for an $\alpha$-rich freeze-out during seed nucleus formation, enabling a large enough neutron-to-seed ratio for the synthesis of heavy $r$-process elements in the escaping ejecta (Sec.~\ref{sec:rprocess}) and powering a short-lived kilonova-like transient (Sec.~\ref{sec:kilonova}).}
    \label{fig:cartoon}
\end{figure*}

We illustrate this process schematically in Fig.~\ref{fig:cartoon}.  We envision that one side of the magnetar magnetosphere undergoes reconnection or strong magnetic field dissipation (panel 1).  The sudden release of energy that accompanies the beginning of the GF produces a high-pressure region above the NS surface comprised of photons and electron/positron pairs (red region in panel 2).  We denote this pressure as $P_{\rm GF}$.  The geometric thickness of this region is of order the initial magnetospheric energy density scaleheight.  We denote the radial thickness of the higher-pressure region as $\Delta R_{\rm GF}$ throughout this paper and we envision that it is of order or less than the NS radius ($\Delta R_{\rm GF}\lesssim R_{\ns}$).  

The high pressure of the GF shell $P_{\rm GF}$ drives two shockwaves.  The first is outwards into the low-density region outside the high-pressure shell.  The second shockwave is into the NS atmosphere (denoted by the teal region in panel 2).  The outwards-propagating shock meets no resistance and expands at ultra-relativistic speeds (outward arrows with $v_{\rm sh}\simeq c$ in panel 3).  Such an ultra-relativistic shock, as it collides and shocks the gaseous environment surrounding the magnetar on much larger scales $\gtrsim 10^{12}$ cm, may produce a coherent radio burst akin to an FRB through the synchrotron maser mechanism (e.g., \citealt{Lyubarsky14,Beloborodov17,Metzger+19}). The inwards-propagating shockwave heats the NS crust while simultaneously encountering the very steep density profile of the NS.  The sudden release of pressure from the outer edge of the high-pressure shell is communicated to smaller radii by a rarefaction wave, which moves inwards and eventually reduces the pressure being applied to the star on the characteristic timescale $\sim \Delta R_{\rm GF}/c_{\rm s} \lesssim 10\mu s$ over which the shell radius appreciably expands, where $c_{\rm s} \simeq c/\sqrt{3}$ is the sound speed of the relativistically-hot plasma.  

Meanwhile, the inwards-propagating shock moves deeper into the NS crust.  The region behind it is separated from the high-pressure shell by a contact discontinuity (teal-red transition in panel 3).  As the shock moves inwards, it encounters regions of the crust with higher and higher pressure, slowing the shock down, and weakening it until its speed eventually drops below that of the rarefaction $c_s = c/\sqrt{3}$.  The shockwave effectively terminates where the initial pressure of the high-pressure region ($P_{\rm{GF}}$ in panel 2) becomes approximately equal to the pressure at some depth within the NS crust $P_{\rm cr}\simeq P_{\rm GF}$.  The shocked NS crust, which is originally pushed inwards by the inward-propagating shock, has a much higher specific energy than before the GF eruption, and now starts to expand outwards, transforming its thermal energy back into kinetic energy (akin to a spring uncoiling or a thermal blastwave).   

During these phases, at least a portion of the magnetosphere becomes open because of the magnetosphere dissipation that initiated the GF.  The open magnetosphere allows this super-heated material to escape (panel 4).  As the flare subsides and the pressure is relieved, decompression of the shock-heated surface layers reconverts a fraction of the deposited thermal energy into bulk kinetic energy, allowing matter to escape (see Sec.~\ref{sec:fiducial}).  

The magnitude of the pressure in the GF determines the depth to which the NS is shocked, as follows from the approximate equality $P_{\rm GF}\simeq P_{\rm cr}$.  While the outer crust may be composed of lighter nuclei like iron with electron fraction $0.464$, $Y_e$ decreases with depth in the NS crust (see Sec.~\ref{sec:crust}).  As we show below, and as can be confirmed analytically, for large enough $P_{\rm GF}$, the shock-heated material should dissociate into free nucleons, leading to the ejection of high-entropy and relatively low-$Y_e$ material, again depending on the initial value of $P_{\rm GF}$.  As it expands, the free nucleons will recombine, first to $\alpha$ particles and, via subsequent $\alpha$ and neutron captures to heavier elements via the $r$-process (Sec.~\ref{sec:rprocess}; as in other contexts; e.g., \citealt{qian1996}).  An optical transient is expected from the expanding ejecta, as in models of kilonovae (Sec.~\ref{sec:kilonova}).

The amount of material unbound $M_{\rm{ej}}$ from the NS crust is easily estimated at order-of-magnitude using several different analytic criteria based on either the total energy of the GF or the total energy density of the shocked crustal material.  For example, an absolute upper limit on $M_{\rm{ej}}$ arises from the \emph{global} criterion that the total energy of the GF exceeds the minimal kinetic energy required to unbind the ejecta from the gravitational well of the NS.  A more stringent, but more realistic {\it local} upper limit on the ejecta mass arises by considering only those layers of the shocked crust which separately achieve positive energy.  Considering that the NS crust starts from rest, this criterion is effectively that the internal energy of the shocked material exceeds the gravitational binding energy.  We expand on these  estimates of the ejected mass in Sec.~\ref{sec:analytic} and compare with our numerical calculations in Sec.~\ref{sec:many_sim}.

The material will only escape if it is not radiative.  For most of the parameter space we explore, by post-processing our numerical results, we find that neutrino losses are of negligible importance to the dynamics of the outflowing ejecta.  For sufficiently large $P_{\rm GF}$, corresponding to extremely powerful GF not yet observed, neutrino cooling may become dynamically important (Sec.~\ref{sec:results}).

\subsection{This Paper}
Although the physical processes that give rise to a GF and its aftermath are undoubtedly a complex multi-dimensional magneto-hydrodynamic event, here we present a simplified model as a preliminary exploration.  We perform a suite of 1D spherically-symmetric special-relativistic hydrodynamic simulations, with a simplified $\Gamma$-law equation of state and ignoring radiative losses.  As described above and sketched in Fig.~\ref{fig:cartoon}, we initialize our simulations by setting up a high-pressure shell of defined width above the NS surface, and investigate the outcome for different values of the pressure $P_{\rm GF}$ and the high-pressure shell width $\Delta R_{\rm GF}$, under the admittedly strong assumption that the effects of magnetar-strength magnetic fields can be neglected after a significant fraction of the magnetic field has been dissipated.  Let us stress again here that the unbound ejecta from the Dec. 2004 GF could not have been spherically symmetric because the observed X-ray emission would have been obscured during the first minutes following the flare due to large scattering optical depth \citep{granot2006}.  Thus, our line of sight had to be along a relatively baryon-poor and radiation-rich portion of the outflow.  Multi-dimensional models to understand the ejecta asymmetry will be the subject of future work. 

This paper is organised as follows.  We start by building a simple analytical NS crust model in Sec.~\ref{sec:crust}.  We describe the setup of our simulations and diagnostics of numerical results in Sec.~\ref{sec:numerical}.  We present our simulation results in Sec.~\ref{sec:results}, starting with an in-depth analysis of the Fiducial model, and then moving on to explore a wider suite of simulations covering a wide range of initial shell pressures and widths ($\sim$ GF energies), comparing our results for the ejecta properties to the analytic estimates.  In Sec.~\ref{sec:discussion} we summarize our results and discuss their consequences for a variety of topics, including $r$-process nucleosynthesis, kilonova-like transients resulting from heavy element ejection, and the environments surrounding repeating FRB sources.  Finally, we show our analytic estimates for mass-weighted ejecta distributions in Appendix~\ref{app:analytic} and the comparison of ejecta distributions for different simulation setups in Appendix~\ref{app:hist_comp}.

\section{Crust Model}\label{sec:crust}

We begin by describing the initial pre-GF radial structure of the NS crust, which we assume to be cold (i.e., non-accreting; though see \citealt{chamel2008} for a more general discussion).  The crustal material resides in its absolute ground state in nuclear equilibrium, i.e. ``cold catalysed material'' (though we note this may not always be a good approximation if multiple GFs occur in rapid succession, ejecting matter faster than the crust can re-establish $\beta-$equilibrium; see Sec.~\ref{sec:FRB} for further discussion).  Below the low-density NS atmosphere, resides the ``outer crust", which starts at densities $\rho \approx 10^{4} \: \rm{g} \: \rm{cm}^{-3}$ and extends to a depth corresponding to the neutron drip line at $\rho_{\rm{ND}} \approx 4.3 \times 10^{11} \: \rm{g} \: \rm{cm}^{-3}$ \citep{ruster2006}.  Below the outer crust, the ``inner crust" extends to nuclear density at $\rho_{\rm{nuc}} = 2.8 \times 10^{14} \: \rm{g} \: \rm{cm}^{-3}$ \citep{chamel2008}.  The total baryonic mass of the outer crust for a typical $1.4 \: \Msun$ NS is of order $10^{28} - 10^{29} \: \rm{g}$ ($10^{-5}$--$10^{-4} \: \Msun$) \citep{pearson2011}.  Considering that the inferred baryonic ejecta from the December 2004 GF was $\lesssim 10^{26} \: \rm{g}$ \citep{granot2006}, we conclude that GFs of even significantly greater energy than those observed thus far, will mainly affect the outer crust.  The equation of state (EOS) of cold catalysed matter for $\rho < \rho_{\rm{ND}}$ should therefore be sufficient to describe those layers of the star affected by the GF-driven shockwave dynamics.

\subsection{Structure of the Pre-Flare Crust}\label{sec:eos}
\begin{figure}
	\includegraphics[width=\columnwidth]{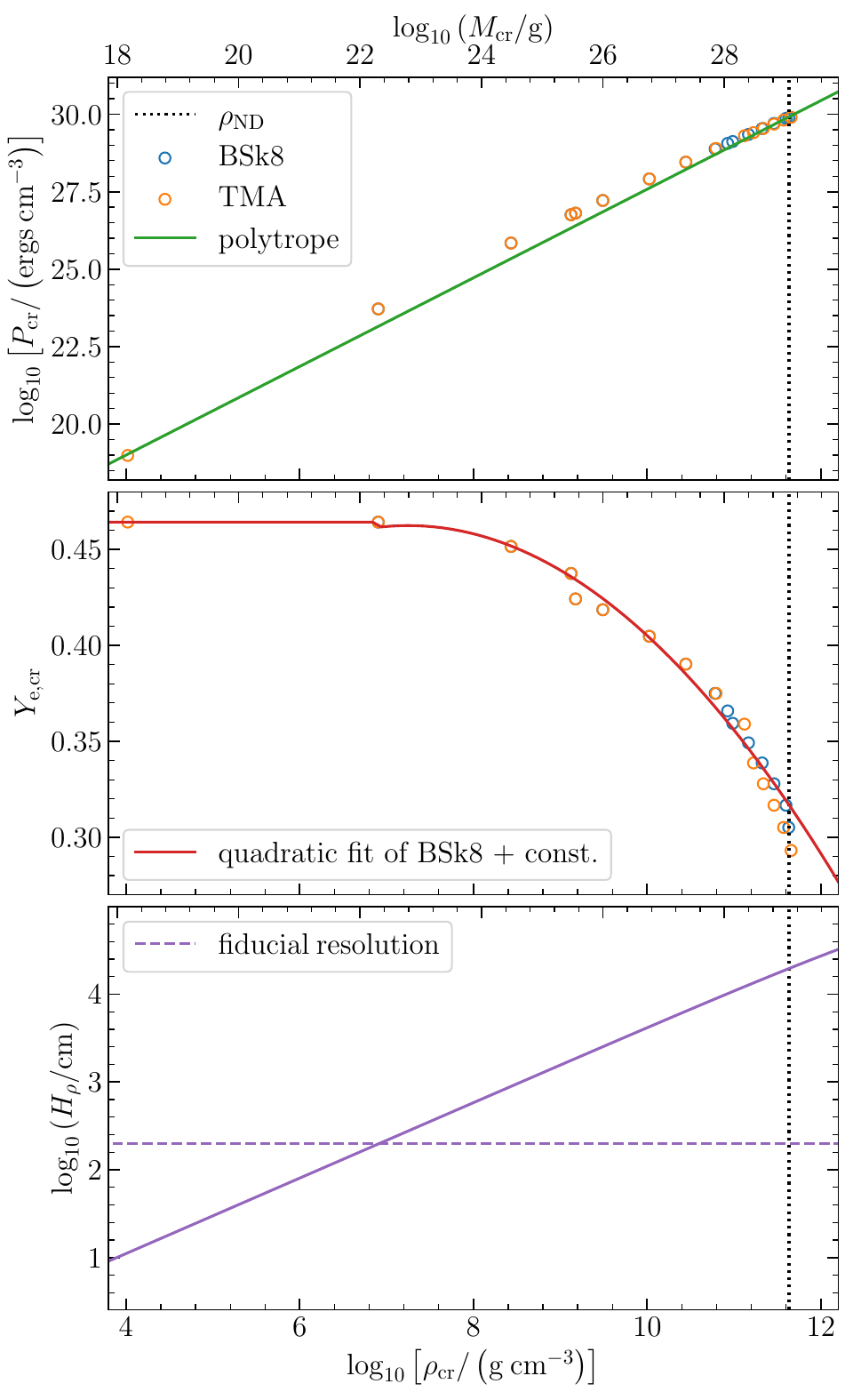}
    \caption{Pressure $P_{\rm cr}$ (top panel), electron fraction $Y_{\rm{e,cr}}$ (middle panel), and density scale-height $H_\rho$ (bottom panel) as functions of density $\rho_{\rm cr}$ in the outer crust of a cold (i.e., non-accreting) NS.  A second horizontal axis along the top of the plot shows the mass depth coordinate $M_{\rm{cr}}$ (equation \ref{eq:M_cr}).  The empty circles indicate the results calculated with the BSk8 (blue; \citealt{samyn2004}) or TMA (orange; \citealt{geng2005}) theoretical nuclear mass tables as explained in the text \citep{ruster2006,byam1971,audi2003}.  The solid lines indicate our approximations, the polytrope (green; equation \ref{eq:polytrope}), the constant $Y_{\rm{e,Fe}}$ $+$ the quadratic fit of $Y_{\rm{e,cr}}(\rho_{\rm cr})$ (red), and the density scale-height $H_\rho$ (purple).  The purple dashed line shows the resolution of our fiducial model (see Table \ref{tab:results}).  The black dotted vertical line indicates the neutron drip line at $\rho_{\rm{ND}}$.}
    \label{fig:P-Y_e-H_rho}
\end{figure}

The EOS of the outer crust of nonaccreting cold NSs was intensively studied by \citet{ruster2006}, who utilized the theory of \citet{byam1971}, experimentally measured atomic masses from \citet{audi2003}, and made different assumptions about the nuclear mass model at densities higher than those accessible to laboratory experiment.  \citet{ruster2006} show that the EOS for $\rho < \rho_{\rm{ND}}$ is well-established, with only small variations in the pressure of a few percent of pressure at a given $\rho$ for different theoretical nuclear mass models.  Hence, we only show their results here for two different theoretical nuclear mass models, namely, the nonrelativistic Skyrme force model BSk8 of \citet{samyn2004} and the relativistic mean field TMA model of \citet{geng2005}, see the top and the middle panel in Fig.~\ref{fig:P-Y_e-H_rho}.  For our purposes we only need the crustal pressure as a function of the crustal density $P_{\rm cr}(\rho_{\rm cr})$ (top panel) and the crustal electron fraction $Y_{\rm{e,cr}}(\rho_{\rm cr})$ (middle panel).

We approximate the pressure of the crust as a polytrope,
\begin{equation}\label{eq:polytrope}
    P_{\rm cr} = P_* \lp \frac{\rho_{\rm cr}}{\rho_*}\rp^{\Gamma_*} = K_* \rho_{\rm cr}^{\Gamma_*},
\end{equation}
where $\log_{10} \left[ P_* / \lp \rm{ergs} \: \rm{cm}^{-3}\rp \right] = 19$, $\log_{10} \left[ \rho_* / \lp \rm{g} \: \rm{cm}^{-3}\rp \right] = 4$, $K_*$ is the polytropic constant and $\Gamma_* = 1.43$ is the approximate effective polytropic exponent (fit to the crust pressure-density relationship).  To approximate the  electron fraction $Y_{\rm{e,cr}}(\rho_{\rm cr})$ we use the constant $Y_{\rm{e,Fe}} = 26/(26+30) \approx 0.464$ for $\log_{10} \left[ \rho_{\rm cr} / \lp \rm{g} \: \rm{cm}^{-3}\rp \right] < 6.90$ and the quadratic fit of the BSk8 model with the least square method for $\log_{10} \left[ \rho_{\rm cr} / \lp \rm{g} \: \rm{cm}^{-3}\rp \right] > 6.90$.  The top panel of Fig.~\ref{fig:P-Y_e-H_rho} shows that the polytrope (\ref{eq:polytrope}) approximates the true pressure $P_{\rm cr}(\rho_{\rm cr})$ reasonably well, while the middle panel shows the constant $Y_{\rm{e,Fe}}$ $+$ the quadratic fit is a good approximation of $Y_{\rm{e,cr}}(\rho_{\rm cr})$ for $\log_{10} \left[ \rho_{\rm cr} / \lp \rm{g} \: \rm{cm}^{-3}\rp \right] \in \langle 4,12\rangle$.

\subsection{Density and Mass Coordinates}
Neglecting general-relativistic corrections, the equation of hydrostatic equilibrium reads
\begin{equation}\label{eq:hydrostat}
    \frac{1}{\rho_{\rm cr}} \frac{\dd P_{\rm cr}}{\dd r} = -g(r) = - G \frac{M_\ns}{r^2},    
\end{equation}
where $r$ is the radius, $g(r)$ the gravitational acceleration, $G$ the gravitational constant and $M_\ns$ is the total NS mass (we neglect the self-gravity of the crust).  Equations (\ref{eq:polytrope}), (\ref{eq:hydrostat}) then define the crust density profile
\begin{equation}\label{eq:rho_prof}
    \rho_{\rm cr} (r) = \left[ 1 + \frac{\Gamma_*-1}{\Gamma_*} \frac{\rho_\surf}{P_\surf} \frac{G M_\ns}{R_\ns} \lp \frac{R_\ns}{r} - 1\rp\right]^{\frac{1}{\Gamma_*-1}} \! \rho_\surf,
\end{equation}
where $P_\surf = K_* \rho_\surf^{\Gamma_*}$, $R_\ns$ is the NS radius, and $\rho_\surf = \rho(R_\ns), P_\surf = P(R_\ns)$ are the surface density and pressure, respectively.  The density scale-height is defined by $H_\rho \equiv - \rho_{\rm cr} / (\dd \rho_{\rm cr}/ \dd r)$, as shown by the solid line in the bottom panel of Fig.~\ref{fig:P-Y_e-H_rho}.

We define an analytical mass coordinate $M_{\rm{cr}}(>r)$ as the mass above radius $r$, i.e.~$M_{\rm{cr}}(R_\ns) \equiv 0$, so that
\begin{equation}\label{eq:M_cr}
    M_{\rm{cr}}(>r) \equiv \int_r^{R_\ns} 4 \pi {r^\prime}^2 \rho_{\rm cr}(r^\prime) \dd r^\prime \approx \frac{4 \pi R_\ns^4}{G M_\ns} P_{\rm cr}(r),
\end{equation}
where we have assumed $g \approx g_\surf \equiv G M_\ns/R_\ns^2$ in the crust and $P_{\rm cr}(r) \gg P_\surf$.  We show $M_{\rm{cr}}$ for $M_\ns = 1.4 \: \Msun$ and $R_\ns = 12 \: \rm{km}$ as a second (top) horizontal axis in Fig.~\ref{fig:P-Y_e-H_rho}.  The total mass of the outer crust is $\sim 10^{29} \: \rm{g}$ when integrated up to the neutron drip density $\rho_{\rm ND}$ (vertical dotted line), consistent with the general-relativistic calculations of \citet{pearson2011}.  The bottom panel of Fig.~\ref{fig:P-Y_e-H_rho} shows that the fiducial resolution of our simulations (Sec.~\ref{sec:numerical}) is sufficient to resolve $H_\rho$ down to mass-depth $M_{\rm cr} \sim 10^{22} \: \rm{g}$ but not the surface layers further out.  We use these pressure, density, and mass profiles (equations \ref{eq:polytrope},\ref{eq:rho_prof},\ref{eq:M_cr}) as initial conditions for our numerical experiments (see Sec.~\ref{sec:sim_setup}) and in our analytic estimates of the ejecta mass (Sec.~\ref{sec:analytic}).

\section{Numerical Approach and Diagnostics}\label{sec:numerical}

We perform 1D spherically symmetric-relativistic hydrodynamics (RHD) simulations of GF shockwaves with the \textsc{Pluto} code \citep{mignone2007}.  We describe the \textsc{Pluto} code and the simulation setup in Sec. \ref{sec:sim_setup} and introduce the quantities needed for diagnostics of the simulation data in Sec. \ref{sec:diagnostics}.

\subsection{Simulation Setup}\label{sec:sim_setup}
\textsc{Pluto} is a multiphysics code designed for the treatment of discontinuous astrophysical flows \citep{mignone2007}.  The discontinuities are dealt with utilizing different Godunov-type high-resolution shock-capturing schemes for integrating a system of conservation laws.  A general system of conservation laws can be written in the following form \citep{mignone2007}
\begin{equation}
    \frac{\partial \boldsymbol{U}}{\partial t} = - \boldsymbol{\nabla} \boldsymbol{\cdot} \boldsymbol{\mathsf{T}} (\boldsymbol{U}) + \boldsymbol{S} (\boldsymbol{U}), 
\end{equation}
where $\boldsymbol{U}$ is a vector of conservative quantities, $\boldsymbol{\mathsf{T}}$ denotes a tensor of fluxes of the components of $\boldsymbol{U}$, and $\boldsymbol{S}$ denotes the source term.  \textsc{Pluto} includes different physics modules.  We use the RHD module.  We use an ideal gas constant-$\Gamma$ EOS \citep{mignone2021}, such that the specific enthalpy obeys,
\begin{equation}\label{eq:ideal_gas}
    h = c^2 + \frac{\Gamma}{\Gamma-1} \frac{P}{\rho}.
\end{equation}
where $P$ is the thermal pressure, $\rho$ is the rest-mass density, and  $\Gamma$ (corresponding to the hot shocked gas) is different from the index $\Gamma_{*}$ which defined the initial cold crust (see below).

We initiate the simulation by setting up a shell of uniform high-pressure $P_{\rm{GF}}$ and width $\Delta R_{\rm{GF}} \lesssim R_\ns$ above the NS surface, with the inner boundary at $r = R_\ns$ and the outer boundary at $r = R_\ns + \Delta R_{\rm{GF}}$.  Insofar that GFs are powered by the large-scale reconfiguration and dissipation of the magnetic field, we motivate the chosen values of the pressure assuming it should roughly equal the pressure of the magnetar's pre-GF magnetic field, i.e.~
\begin{equation}\label{eq:P_shock}
    P_{\rm{GF}} \approx \frac{B^2}{8 \pi} \simeq 4\times 10^{28}\:{\rm ergs\:cm^{-3}}\left(\frac{B}{10^{15}\:{\rm G}}\right)^{2},
\end{equation}
where $B$ is the surface magnetic field strength and need not be identical to the polar magnetic field strength inferred from vacuum dipole spindown.  We do not self-consistently include the effects of the magnetic field itself in our 1D simulations; rather, we implicitly assume that following a strong magnetic dissipation event the dynamical effects of the magnetic field can be neglected to first order across at least a portion of the NS surface, with material thus able to freely escape along the torn open magnetic field lines \citep{thompson1995} after the magnetic field reconfiguration (Fig.~\ref{fig:cartoon}; we discuss the limitations of this assumption in Sec.~\ref{sec:caveats}).  

The energy in the high-pressure GF shell is given by,
\begin{equation}
E_{\rm GF} \simeq \Delta \Omega R_{\rm NS}^{2}\Delta R_{\rm{GF}}(3P_{\rm{GF}}) \simeq 5\times 10^{48}\:{\rm ergs}\:P_{\rm{GF},30} \Delta R_{\rm GF,1},
\label{eq:E_GF}
\end{equation}
where $P_{\rm{GF},30} = P_{\rm{GF}} / (10^{30} \: \rm{ergs} \: \rm{cm}^{-3})$, $\Delta R_{\rm GF,1} = \Delta R_{\rm GF} / (1 \: \rm{km})$, $\Delta \Omega$ is the solid angle subtended by the shell (Fig.~\ref{fig:cartoon}), and the final line assumes $\Delta \Omega = 4\pi$ and $R_{\ns} = 12$ km.  For these parameters, the energy scales $E \sim 10^{44}-10^{47}$ ergs of Galactic magnetar GF correspond to $P_{\rm{GF}} \sim 10^{25}-10^{28}$ ergs cm$^{-3}$ for $\Delta R_{\rm GF}=1\:{\rm km}$.  

We assume a NS of mass $M_\ns = 1.4 \: \Msun$.  As our default assumption, we place the inner boundary of the simulation region at $R_{\rm{in}} = 10 \: \rm{km}$, the NS surface at $R_{\rm{NS}} = 12 \: \rm{km}$, and the outer boundary at $R_{\rm{out}} = 20 000 \: \rm{km}$.  The inner boundary condition is reflective, while the outer boundary employs an outflow condition.  The initial density profile inside the star ($r < R_\ns$) is given by our assumed crustal model (equation \ref{eq:rho_prof}), which for numerical purposes we attach onto a power-law profile $\rho(r) = \rho_\surf (r / R_\ns)^{-3}$ above the surface $r > R_\ns$ (containing negligible mass relative to the crust).  The initial pressure profile follows equations (\ref{eq:polytrope},\ref{eq:rho_prof}), for $r < R_\ns$, we assume roughly constant pressure in the high-pressure shell $P = P_{\rm{GF}}$, for $r \in \langle R_\ns, R_\ns + \Delta R_{\rm{GF}}\rangle$, and the same power-law as for the density profile in the outer region $P(r) = P_\surf [\rho(r)/\rho_\surf]$, for $r > R_\ns + \Delta R_{\rm{GF}}$.  We set the minimum density and pressure in the initial profiles to be $\rho_{\rm{floor}} \simeq 2 \times 10^{-24} \rho_{\rm ND}$, $P_{\rm{floor}} \simeq 1 \times 10^{-24} P_{\rm cr} (\rho_{\rm ND})$ to avoid numerical instabilities.

We take $\Gamma = 4/3$ in our simulations (cf. $\Gamma_* = 1.43$, equation \ref{eq:polytrope}), motivated by the fact that the pressure and energy density of the high-pressure shell and shocked crustal material is dominated by radiation and relativistic non-degenerate electron/positron pairs.  By adopting a $\Gamma-$law EOS, we also neglect the energy released by the shock-dissociation of alpha particles into free nucleons and their eventual recombination in the outflow (Sec.~\ref{sec:nucleosynthesis}); this is a reasonable approximation because the energy released $\simeq $ 7 MeV per nucleon is modest compared to the asymptotic specific kinetic energy of the unbound ejecta, which is typically comparable to the NS gravitational binding energy $\approx 150$ MeV per nucleon.  For our default simulation B15.0\_1km we set $\rho_\surf = 1 \: \rm{g} \: \rm{cm}^{-3}$, $P_{\rm{GF}} = 3.98 \times 10^{28} \: \rm{ergs} \: \rm{cm}^{-3}$ (corresponding to $B = 10^{15} \: \rm{G}$), and $\Delta R_{\rm{GF}} = 1 \: \rm{km}$.  However, we also run a suite of simulations spanning a wide range of $P_{\rm{GF}} \sim 10^{25} - 10^{32}$ ergs cm$^{-3}$ and $\Delta R_{\rm GF} = 0.1 - 10$ km that corresponds to $E_{\rm GF} \sim 10^{43} - 10^{51} \: \rm{ergs}$.

As our default assumption, we employ a uniform radial grid of $N_{\rm{u}} = 1500$ points between $R_{\rm{in}}$ and $R_{\rm{u-s}} = 13 \: \rm{km}$ and stretched grid of $N_{\rm{s}} = 10000$ points up to the outer boundary at $R_{\rm{out}}$.  Our default uniform resolution of $2 \: \rm{m}$ near the NS surface allows us to resolve the density scale height interior to a mass coordinate of $\sim 10^{22} \: \rm{g}$, as shown in the bottom panel of Fig.~\ref{fig:P-Y_e-H_rho}.  The length of the $n$-th stretched grid element is $q^n_{\rm{s}} \Delta r_{\rm{u}}$, where $q_{\rm{s}}$ is the stretching ratio and $\Delta r_{\rm{u}}$ is the length of a uniform grid element.  By default, we use a piecewise total variation diminishing (TVD) linear reconstruction, 2nd-order TVD Runge-Kutta time stepping, CFL condition of $0.5$, and a simple TVD Lax-Friedrichs Riemann solver \citep{mignone2021}.
However, as we shall discuss in Sec.~\ref{sec:convergence}, we also explore the sensitivity of our results to the resolution, reconstruction technique (from linear to parabolic; \citealt{mignone2014}) and Riemann solver (TVD Lax-Friedrichs to two-shock HLLC; \citealt{mignone2006}), finding moderate quantitative differences in key quantities such as ejecta mass and entropy distribution relative to those obtained for the default setup.  Let us also mention that a test simulation performed without a high-pressure shell ($P_{\rm GF} = 0$) gives rise to no ejecta.  Each simulation is run for approximately one light-crossing time of the domain, i.e. $67 \: \rm{ms}$, which we find is sufficient for convergence of the unbound ejecta properties.

\subsection{Diagnostics and Analysis}\label{sec:diagnostics}
To analyze the effects of neutrino cooling and nucleosynthesis, we must extract the fluid temperature $T$ from our simulation data.  Under the conditions of high temperatures $kT \gg m_e c^{2} \approx 0.5$ MeV and high entropies which characterize the shock-heated NS crust material, electrons/positrons are relativistic and form a non-degenerate ideal gas.  The total pressure is therefore comprised of baryons and radiation (photons, electron, positrons) in thermal equilibrium, for which:
\begin{equation}\label{eq:T_approx}
    T \approx \min \Biggl\{ \frac{\mu m_{\rm{u}}}{k} \frac{P}{\rho}, \lp \frac{12 P}{11a}\rp^{1/4} \Biggr\},
\end{equation}
where $m_{\rm{u}}$ is the atomic mass unit, $k$ is the Boltzmann constant, $a$ is the radiation constant, and $\mu = 1$ is the mean molecular weight of the free nucleons.  The entropy per baryon in relativistic particles in units of $k$ can be approximated as \citep{qian1996}
\begin{equation}\label{eq:S_b}
    S_{\rm{b}} \simeq 5.21 T_{\rm{MeV}}^3 \rho^{-1}_8,
\end{equation}
where $T_{\rm MeV} \equiv kT/{\rm MeV}$ and $\rho_8 = \rho / (10^8 \: \rm{g} \: \rm{cm}^{-3})$.  Radiation pressure dominates over ion gas pressure for $S_{\rm{b}} \gg 1.$ 

The importance of degeneracy effects on the populations of electrons/positrons can be assessed by the ratio $\eta_{\rm e} \equiv \mu_{\rm e} / (k T)$, where $\mu_{\rm e}$ is the electron chemical potential parameter.  Using the approximate expression (\citealt{qian1996}; their Eq.~6),
\begin{equation}\label{eq:eta_e}
    \eta_{\rm e} \lp 1 + \frac{\eta_{\rm e}^2}{\pi^2}\rp = 1.388 \frac{Y_{\rm e}\rho_8}{T_{\rm MeV}^{3}},
\end{equation}
we shall find $\eta_{\rm e} \ll 1$ (equivalently, $S_{\rm b} \gg 2.4$) for the shocked layers ultimately unbound from the star.  This illustrates that degeneracy effects on the EOS and pair-capture weak interaction rates can be neglected to first order. 

We now introduce several of the key diagnostic quantities associated with neutrino cooling (Sec.~\ref{sec:cooling}), nucleosynthesis (Sec.~\ref{sec:nucleosynthesis}), and for estimating the unbound ejecta layers (Sec.~\ref{sec:mat_ex}).

\subsubsection{Neutrino Cooling and Leptonization}\label{sec:cooling}
Although the photon optical depths are sufficiently high surrounding the star during the post-flare evolution that photons are essentially trapped in the fluid, the shocked layers are transparent to neutrinos.  As we do not include radiative cooling in our simulations, we must therefore check that neutrino cooling has a negligible impact on the dynamics of the system.  An approximate condition for this to be justified is that the expansion timescale of the shocked gas, $t_{\rm{exp}}$, be shorter than the neutrino cooling time-scale, $t_{\nu}$.  We estimate the local expansion time-scale as
\begin{equation}\label{eq:t_exp}
    t_{\rm{exp}} = \frac{r-R_\ns}{\left| v \right|},
\end{equation}
where we use the absolute value $\left| v\right|$ because the velocity can become negative close to the NS surface in our simulations.  An alternative but closely related global dynamical time-scale, $t_{\rm{dyn}}$, related to the unbound ejecta layers, will be defined in Sec.~\ref{sec:mat_ex}.

For the neutrino cooling timescale, we estimate
\begin{equation}\label{eq:t_nu}
  t_{\nu} = \frac{3P}{Q_{\nu}},
\end{equation}
where $Q_{\nu} = Q_{\rm pp} + Q_{\rm cc}$ is the specific neutrino cooling rate.  For the latter, we consider the two most relevant processes: pair-production, $\nu_i \bar{\nu}_i \leftrightarrow \rm{e}^+ \rm{e}^-$, and charged-current electron/positron capture onto nucleons, $\nu_{\rm{e}} \rm{n} \leftrightarrow \rm{e}^- \rm{p}, \bar{\nu}_{\rm{e}} \rm{p} \leftrightarrow \rm{e}^+ \rm{n}$. 

The combined neutrino cooling rate per unit volume for pair-production by $\nu_{\rm{e}} \bar{\nu}_{\rm{e}}$, $\nu_\mu \bar{\nu}_\mu$, and $\nu_\tau \bar{\nu}_\tau$ is approximately given by \citep[e.g.,][]{thompson2002}
\begin{equation}\label{eq:Q_pp}
    Q_{\rm{pp}} \approx 1.4 \times 10^{25} \: \rm{ergs} \: \rm{cm}^{-3} \: \rm{s}^{-1} \: \mathit{T}_{\rm{MeV}}^9.
\end{equation}
Likewise, the cooling rate associated with the charged-current processes is \citep[e.g.,][]{thompson2001}
\begin{equation}\label{eq:Q_cc}
    Q_{\rm{cc}} \approx 2.0 \times 10^{26} \: \rm{ergs} \: \rm{cm}^{-3} \: \rm{s}^{-1} \: \mathit{T}_{\rm{MeV}}^6 \rho_{8}.
\end{equation}
We also define the total energy radiated away in neutrinos:
\begin{equation}\label{eq:E_nu}
    E_{\nu}(t) = \int_0^t \int_{R_{\ns}}^{R_{\rm{out}}} 4 \pi r^2 Q_{\nu} \dd r \dd t^\prime.
\end{equation}
Except for the most powerful magnetar GF, we shall find $t_{\nu} \gg t_{\rm dyn}$ is satisfied in the layers ultimately unbound from the star; this indicates that (1) neutrino losses can be neglected on the ejecta dynamics; (2) after being shock-heated, the entropy $S_{\rm b}$ in the unbound layers remains largely conserved as they decompress and undergo nucleosynthesis.

Although the material ultimately ejected from the star originates from the neutron-rich crust ($Y_e < 0.5)$, weak interactions in the hot shocked ejecta can in principle also change $Y_e$ from these initial values.  Neglecting neutrino absorption reactions, the electron fraction evolves as (e.g., \citealt{qian1996})
\begin{equation}
\frac{dY_e}{dt} = - \lambda_{\rm e-p}Y_e + \lambda_{\rm e+n}(1-Y_e),
\end{equation}
where $\lambda_{\rm e-p}$ is the rate of $\rm{e}^{-}\rm{p} \rightarrow \rm{n} \nu_{\rm{e}}$ and $\lambda_{\rm e+n}$ is the rate of $\rm{e}^{+}\rm{n} \rightarrow \rm{p} \bar{\nu}_{\rm{e}}$.

The electron fraction therefore evolves on a timescale \citep[e.g.,][]{Beloborodov03},
\begin{equation}\label{eq:t_Y_e}
t_{Y_e} \equiv \frac{Y_e}{dY_e/dt} = \frac{Y_e}{1-2Y_e}\frac{1}{\lambda} \simeq 2.3\:{\rm s} \: \frac{Y_e}{1-2Y_e}T_{\rm MeV}^{-5}\:
\end{equation}
where the final line assumes the high-temperature non-degenerate limit, $\lambda_{\rm e+n} \approx \lambda_{\rm e-p} \approx \lambda \propto T^{5}$.  We shall find that $t_{Y_e} \gg t_{\rm exp}$ is always satisfied in the unbound ejecta layers; this indicates that weak interactions are slow and hence $Y_e$ of the ejecta layers will retain their original values from the NS crust (Fig.~\ref{fig:P-Y_e-H_rho}).

\subsubsection{Nucleosynthesis}\label{sec:nucleosynthesis}
We first must address whether the shocked crustal material will be dissociated into free nucleons (protons and neutrons).  The free nucleon mass fraction can be approximated by the following expression: \citep{woosley1992}
\begin{equation}\label{eq:X_N}
    X_{\rm{N}} = \min \Biggl\{ 1, 828 \frac{T_{\rm{MeV}}^{9/8}}{\rho_8^{3/4}} \exp \lp -\frac{7.074}{T_{\rm{MeV}}}\rp \Biggr\}.
\end{equation}
As we shall show, most of the crustal layers ultimately unbound from the star are shock heated to sufficiently high temperatures to be initially dissociated ($X_{\rm{N}} \simeq 1$).

As these shocked layers cool and decompress adiabatically away from the star, heavy element nucleosynthesis can in principle occur.  In particular, once the temperature drops below $\sim 0.5-1$ MeV, free nucleons will recombine into alpha particles, which themselves can undergo a neutron-aided version of the triple alpha process to create $^{12}$C and ultimately (after additional $\alpha$ captures) heavier ``seed nuclei'' \citep{woosley1992}.  When the ejecta is neutron-rich ($Y_e < 0.5$), any remaining free neutrons can then capture onto these seed nuclei, creating $r$-process elements (e.g., \citealt{hoffman1997}).  Neutron captures proceed to create nuclei up to a maximum mass number determined by the ratio of neutrons to seed nuclei.  

For the moderately neutron-rich composition $Y_e \approx 0.4-0.5$ of the unbound ejecta in our simulations, heavy $r$-process production beyond the second or third peak is only possible by suppressing the formation of seed nuclei (so-called ``alpha-rich freeze-out''), as occurs for low densities (high entropy) or rapid expansion timescale.  In particular, whether the $r$-process proceeds up to the third peak (nuclear mass number $A \sim 195$) can be characterized by a single quantity \citep{hoffman1997,thompson2018}
\begin{equation}\label{eq:zeta}
    \zeta \equiv \frac{S_{\rm{b}}^3}{Y^3_{\rm{e}} t_{\alpha,0}}, \quad \zeta_{\rm{crit}} \approx 8 \times 10^9,
\end{equation}
where $t_{\alpha,0} = t_{\rm{exp}}(t, R_\alpha) / (1 \: \rm{s})$ is the expansion time (equation \ref{eq:t_exp}) measured at the radius $R_\alpha$ where seed nuclei begin to form (typically, $T_{\rm MeV} \approx T_{\alpha,\rm MeV} = 0.5$).  We define the $\alpha$-particle formation radius according to 
\begin{equation}\label{eq:R_alpha}
    R_\alpha (t) \equiv \min_{T_{\rm MeV}(t,r) \leq 0.5} \{ r\},
\end{equation}
where we take the minimum because the temperature profile is not monotonic in radius.  To approximate the expansion time $t_{\alpha,0}$ of the ejecta at a given radius $r$, we assume homologous expansion, i.e.~$\rho \propto r^{-3}$.  Further, assuming the entropy per baryon $S_{\rm b}$ (equation \ref{eq:S_b}) in the ejecta remains constant during its expansion, we have $T \propto r^{-1}$.  This allows us to estimate $t_{\alpha,0}$ as follows
\begin{equation}\label{eq:t_exp,alpha}
    t_{\alpha,0} (t,r) = \frac{R_\alpha (t) - R_\ns}{r - R_\ns} t_{\rm{exp},0} (t,r) \approx \frac{T(t,r)}{T_\alpha} t_{\rm{exp},0} (t,r),
\end{equation}
where $t_{\rm{exp},0} = t_{\rm exp} / (1 \: \rm{s})$ and we have assumed $R_\alpha \gg R_\ns$, $r \gg R_\ns$.  If $\zeta > \zeta_{\rm{crit}}$, then the third $r$-process peak is achieved, while if instead $\zeta < \zeta_{\rm{crit}}$ the $r$-process terminates at lower $A \ll 195$.  The general trend of the analytical expression agrees well with the numerical simulations and thus serves as a useful guideline for the third $r$-process peak detection \citep{thompson2018}.

\subsubsection{Unbound Ejecta}\label{sec:mat_ex}
In analogy to equation (\ref{eq:M_cr}) we can define a time-dependent mass coordinate $M(t,>r)$ as the mass above a given radius $r$, this time calculated directly from the simulation data,
\begin{equation}\label{eq:M_num}
    M(t,>r) \equiv \int_r^{R_{\rm{out}}} 4 \pi {r^\prime}^2 \rho(t,r^\prime) \dd r^\prime,
\end{equation}
where again $M(t, >R_{\rm{out}}) = 0$.  The time-dependent ejecta mass $M_{\text{ej},r} (t, >r)$ above a given radius $r$ can thus be defined by
\begin{equation}\label{eq:M_ej,r}
    M_{\text{ej},r}(t,>r) \equiv \int_{r}^{R_{\rm{out}}} \left. 4 \pi {r^\prime}^2 \rho(t,r^\prime) \dd r^\prime \right\vert_{e(t,r^\prime) > 0},
\end{equation}
where
\begin{equation}\label{eq:en_dens}
    e = \frac{1}{2} v^2 + \epsilon + \phi = \frac{1}{2} v^2 + \frac{1}{\Gamma-1} \frac{P}{\rho} - G \frac{M_\ns}{r},
\end{equation}
is the total energy density, comprised of kinetic, internal, and gravitational potential components, respectively.  Mass shells with $e > 0$ are considered to form part of the unbound ejecta.  Insofar that our simulations are only run slightly longer than the light-crossing time over the domain, we are justified to neglect any mass leaving the outer boundary of the simulation domain.  We define the minimum radius $R_{\rm ej,min} (t)$ of the unbound ejecta according to \begin{equation}\label{eq:R_ej,min}
    R_{\rm{ej,min}} (t) = \min_{e(t,r) > 0} \{ r\}.
\end{equation}
Thus, the total ejecta mass at a given time $t$ becomes
\begin{equation}\label{eq:M_ej}
    M_{\rm ej} (t) = M_{\text{ej},r} (t, >R_{\rm in}) = M_{\text{ej},r} (t, >R_{\rm{ej,min}} (t)).
\end{equation}

We define a {\it global} dynamical timescale of the ejecta as
\begin{equation}\label{eq:t_dyn}
    t_{\rm dyn} = - \lp \frac{1}{\rho_{0.5}} \frac{\dd \rho_{0.5}}{\dd t}\rp^{-1},
\end{equation}
where $\rho_{0.5}(t) = \rho(t,R_{\rm{ej},0.5}(t))$ and $R_{\rm{ej},0.5}(t)$ is the half-mass radius defined by
\begin{equation}\label{eq:R_ej,0.5}
    M_{\text{ej},r}(t, >R_{\rm{ej},0.5}(t)) = 0.5 M_{\rm{ej}}(t).
\end{equation}
This estimate is meant to approximate, in a single quantity, the characteristic expansion time experienced by the bulk of the unbound ejecta (as distinct from the {\it local} expansion timescale defined in equation~\ref{eq:t_exp}).

Finally, to estimate the electron fraction of the ejected material, $Y_{\rm{e}}$ (e.g., as enters equation \ref{eq:zeta}) we use the mapping $Y_{\rm{e}}(t,r) = Y_{\rm e,cr}(\rho(M(t,>r)))$ (see Fig.~\ref{fig:P-Y_e-H_rho}), where $\rho(M(t,>r))$ is obtained by inverting equations (\ref{eq:polytrope}) and (\ref{eq:M_cr}), respectively, i.e.~$\rho = \rho_{\rm cr}(P_{\rm cr}(M_{\rm{cr}}))$.

\section{Results}\label{sec:results}

Table \ref{tab:results} summarizes our suite of simulations and several key properties of the ejecta, while Table~\ref{tab:convergence} shows convergence tests performed for a single model.  We begin in Sec.~\ref{sec:fiducial} by describing in detail the Fiducial model with $B = 10^{16}$ G ($P_{\rm GF} = 10^{30.6}$ ergs cm$^{-3}$) and $\Delta R_{\rm GF} = 1 \: \rm{km}$.  In Sec.~\ref{sec:convergence} we perform convergence tests which assess the sensitivity of our results to things like the the numerical scheme and adopted resolution. Then in Sec.~\ref{sec:analytic} we derive analytic constraints on the unbound ejecta mass, following the physical picture outlined in Sec.~\ref{sec:overview}. In Sec.~\ref{sec:many_sim} we describe the broader simulation results covering a wide range of shock strengths (different $P_{\rm{GF}}$) and shell-thicknesses ($\Delta R_{\rm{GF}}$) and describe the summary trends and compare them to the analytic estimates.

\begin{table*}
    \centering
    \caption{Simulation results.  The superscript $^*$ denotes simulations artificially terminated before $67 \: \rm{ms}$.  The subscript \_two-sh indicates simulations in which the Riemann solver is changed from TVD Lax-Friedrichs to two-shock HLLC \citep{mignone2021}.  Note the short-hand notation $\log_{10} \equiv \lg$ and that $\overline{\text{quantity}_{\rm ej}}$ denotes a mass-weighted average of the unbound ejecta as defined by equation (\ref{eq:M_ej}) as measured at the end of the simulation.}\label{tab:results}
    \begin{tabular}{lcccccc @{\extracolsep{10pt}} ccccc}
        \hline
        \hline
        Model name & \multicolumn{6}{c}{Model parameters} & \multicolumn{5}{c}{Ejecta Properties}\\
        \cline{2-7}
        \cline{8-12}
        \rule{-0.6ex}{3ex}
         & $\lg B$ & $\lg P_{\rm GF}$ & $\Delta R_{\rm{GF}}$ & $\lg\rho_{\rm{surf}}$ & $N_{\rm{u}}$ & $N_{\rm s}$ & $\lg M_{\rm{ej}}$ & $\overline{v_{\rm{ej}}}$ & $\overline{Y_{\rm{e,ej}}}$ & $\overline{\lg S_{\rm{b,ej}}}$ & $\overline{\lg\zeta_{\rm{ej}}}$ \\
         \rule{-0.6ex}{3ex}
         & [G] & $[\rm{ergs} \: \rm{cm}^{-3}]$ & [km] & [g] & & & [g] & [$c$] & & &  \\
        \hline
        Fiducial                     & 16.0 & 30.60 &  1 &  2 & 1500 & 10000 &     25.50&      0.28&      0.44&      2.21&      11.4\\
        Fiducial\_0.1km              & -    & -     & 0.1&  - & -    & -     &     25.52&      0.33&      0.44&      2.08&      11.1\\
        Fiducial\_1km\_2x            & -    & -     &  1 &  - & 3000 & -     &     25.45&      0.29&      0.44&      2.21&      11.4\\
        Fiducial\_1km\_0.5x\_two-sh  & -    & -     &  - &  - & 1500 & 5000  &     25.68&      0.33&      0.44&      2.28&      11.6\\
        Fiducial\_2km                & -    & -     &  2 &  - & -    & 10000 &     25.61&      0.30&      0.44&      2.15&      11.2\\
        Fiducial\_10km               & -    & -     & 10 &  - & -    & -     &     26.34&      0.29&      0.43&      1.81&      10.1\\
        \rule{-0.6ex}{3ex}
        B13.5\_0.1km                 & 13.5 & 25.60 & 0.1& -3 & -    & -     &     17.38&      0.23&      0.46&      4.22&      19.3\\
        B13.5\_1km                   & -    & -     &  1 &  - & -    & -     &     18.22&      0.19&      0.46&      4.01&      18.4\\
        B13.5\_1km\_2x               & -    & -     &  - &  - & 3000 & -     &     18.66&      0.15&      0.46&      3.91&      17.8\\
        B13.5\_1km\_0.5x\_two-sh     & -    & -     &  - &  - & 1500 & 5000  &     18.36&      0.20&      0.46&      4.00&      18.3\\
        B13.5\_2km$^*$               & -    & -     &  2 &  - & -    & 10000 &     18.50&      0.21&      0.46&      4.03&      18.3\\
        B13.5\_10km$^*$              & -    & -     & 10 &  - & -    & -     &     18.17&      0.19&      0.46&      4.20&      18.8\\
        \rule{-0.6ex}{3ex}
        B14.0\_0.1km                 & 14.0 & 26.60 & 0.1& -2 & -    & -     &     19.21&      0.26&      0.46&      3.73&      17.4\\
        B14.0\_1km                   & -    & -     &  1 &  - & -    & -     &     19.86&      0.18&      0.46&      3.54&      16.5\\
        B14.0\_1km\_2x               & -    & -     &  - &  - & 3000 & -     &     19.77&      0.17&      0.46&      3.56&      16.6\\
        B14.0\_1km\_0.5x\_two-sh     & -    & -     &  - &  - & 1500 & 5000  &     20.16&      0.55&      0.46&      4.28&      18.8\\
        B14.0\_2km$^*$               & -    & -     &  2 &  - & -    & 10000 &     20.29&      0.17&      0.46&      3.52&      16.3\\
        B14.0\_10km                  & -    & -     & 10 &  - & -    & -     &     20.68&      0.10&      0.46&      3.47&      15.7\\
        \rule{-0.6ex}{3ex}
        B14.5\_0.1km                 & 14.5 & 27.60 & 0.1& -1 & -    & -     &     20.66&      0.26&      0.46&      3.42&      16.1\\    
        B14.5\_1km                   & -    & -     &  1 &  - & -    & -     &     21.14&      0.25&      0.46&      3.22&      15.4\\
        B14.5\_1km\_2x               & -    & -     &  - &  - & 3000 & -     &     21.11&      0.22&      0.46&      3.22&      15.4\\
        B14.5\_1km\_0.5x\_two-sh     & -    & -     &  - &  - & 1500 & 5000  &     21.48&      0.35&      0.46&      3.38&      15.9\\
        B14.5\_2km                   & -    & -     &  2 &  - & -    & 10000 &     21.42&      0.19&      0.46&      3.16&      15.0\\
        B14.5\_10km                  & -    & -     & 10 &  - & -    & -     &     23.08&      0.04&      0.46&      3.06&      13.6\\
        \rule{-0.6ex}{3ex}
        B15.0\_0.1km                 & 15.0 & 28.60 & 0.1&  0 & -    & -     &     22.30&      0.30&      0.46&      2.93&      14.3\\
        B15.0\_1km                   & -    & -     &  1 &  - & -    & -     &     22.49&      0.25&      0.46&      2.94&      14.2\\
        B15.0\_1km\_2x               & -    & -     &  - &  - & 3000 & -     &     22.50&      0.26&      0.46&      2.91&      14.2\\
        B15.0\_1km\_0.5x\_two-sh     & -    & -     &  - &  - & 1500 & 5000  &     22.98&      0.31&      0.46&      2.91&      14.1\\
        B15.0\_2km                   & -    & -     &  2 &  - & -    & 10000 &     22.80&      0.29&      0.46&      2.79&      13.8\\
        B15.0\_10km                  & -    & -     & 10 &  - & -    & -     &     24.94&      0.10&      0.45&      2.70&      12.5\\
        \rule{-0.6ex}{3ex}
        B15.5\_0.1km                 & 15.5 & 29.60 & 0.1&  1 & -    & -     &     24.04&      0.32&      0.46&      2.45&      12.5\\
        B15.5\_1km                   & -    & -     &  1 &  - & -    & -     &     24.00&      0.31&      0.46&      2.59&      12.9\\
        B15.5\_1km\_2x               & -    & -     &  - &  - & 3000 & -     &     23.98&      0.28&      0.46&      2.56&      12.8\\
        B15.5\_1km\_0.5x\_two-sh     & -    & -     &  - &  - & 1500 & 5000  &     24.35&      0.33&      0.46&      2.65&      13.0\\
        B15.5\_2km                   & -    & -     &  2 &  - & -    & 10000 &     24.34&      0.24&      0.46&      2.43&      12.2\\
        B15.5\_10km                  & -    & -     & 10 &  - & -    & -     &     26.46&      0.11&      0.42&      2.21&      10.7\\
        \rule{-0.6ex}{3ex}
        B16.5\_0.1km                 & 16.5 & 31.60 & 0.1&  3 & -    & -     &     26.87&      0.34&      0.41&      1.73&       9.8\\
        B16.5\_1km                   & -    & -     &  1 &  - & -    & -     &     26.84&      0.34&      0.41&      1.89&      10.3\\
        B16.5\_1km\_2x               & -    & -     &  - &  - & 3000 & -     &     26.68&      0.33&      0.42&      1.92&      10.4\\
        B16.5\_1km\_0.5x\_two-sh     & -    & -     &  - &  - & 1500 & 5000  &     26.88&      0.36&      0.41&      1.96&      10.5\\
        B16.5\_2km                   & -    & -     &  2 &  - & -    & 10000 &     27.23&      0.31&      0.40&      1.66&       9.5\\
        B16.5\_10km                  & -    & -     & 10 &  - & -    & -     &     26.78&      0.32&      0.41&      1.92&      10.3\\
        \hline 
        \hline 
    \end{tabular}
\end{table*}

\begin{table*}
    \centering
    \caption{Convergence tests of the model B15.0\_1km with $B = 10^{15}$ G ($P_{\rm GF} = 10^{28.6}$ ergs cm$^{-3}$) and $\Delta R_{\rm GF} = 1 \: \rm{km}$.  The default simulation setup is described in Sec.~\ref{sec:sim_setup}.  The notation of the ejecta properties is the same as in Table~\ref{tab:results}.  Explanations of the changes to the default setup: $^1$we change the NS surface density $\rho_\surf$ from $1 \: \rm{g}$ to $10^2 \: \rm{g}$, $^2$we change the scaling of the initial density (pressure) profile outside of the NS (high-pressure shell) from $\propto r^{-3}$ to $\propto r^{-2}$, $^3$we change the inner boundary $R_{\rm{in}}$ from $10 \: \rm{km}$ to $11 \: \rm{km}$, $^4$we change the outer boundary $R_{\rm{out}}$ from $20000 \: \rm{km}$ to $10000 \: \rm{km}$ and the simulation duration from $66.7 \: \rm{ms}$ to $33.3 \: \rm{ms}$, $^5$we change the values of minimum density $\rho_{\rm floor}/\rho_{\rm ND}$ (pressure $P_{\rm floor}/P_{\rm cr} (\rho_{\rm ND})$) in the initial profile from $2 \times 10^{-24}$ ($1 \times 10^{-24}$) to $2 \times 10^{-22}$ ($1 \times 10^{-22}$), $^6$we change the $\Gamma$-constant from $4/3$ to that of the pre-flare NS crust $\Gamma = \Gamma_* = 1.43$ (Sec.~\ref{sec:eos}), $^7$we switch from 2nd-order to 3rd-order TVD Runge-Kutta time stepping \citep{mignone2021}, $^8$we switch from TVD Lax-Friedrichs to two-shock HLLC Riemann solver \citep{mignone2021}, $^9$we switch from piecewise TVD linear to parabolic reconstruction scheme \citep{mignone2021}.}\label{tab:convergence}
    \begin{tabular}{lccc @{\extracolsep{10pt}} ccccc}
        \hline
        \hline
        Model name & \multicolumn{3}{c}{Model parameters} & \multicolumn{5}{c}{Ejecta Properties}\\
        \cline{2-4}
        \cline{5-9}
        \rule{-0.6ex}{3ex}
         & $N_{\rm{u}}$ & $N_{\rm s}$ & Targeted change & $\lg M_{\rm{ej}}$ & $\overline{v_{\rm{ej}}}$ & $\overline{Y_{\rm{e,ej}}}$ & $\overline{\lg S_{\rm{b,ej}}}$ & $\overline{\lg\zeta_{\rm{ej}}}$ \\
        \rule{-0.6ex}{3ex}
         &  &  &  & [g] & [$c$] & & & \\
        \hline
        B15.0\_1km                      & 1500 & 10000 & default simulation                                           &     22.49&      0.25&      0.46&      2.94&      14.2\\ 
        B15.0\_1km\_2x                  & 3000 & -     & $N_{\rm{u}} \rightarrow 3000$                                &     22.50&      0.26&      0.46&      2.91&      14.2\\ 
        B15.0\_1km\_0.5x                & 1500 & 5000  & $N_{\rm{s}} \rightarrow 5000$                                &     22.50&      0.25&      0.46&      2.93&      14.2\\ 
        B15.0\_1km\_0.5-0.5x            &  750 & -     & $N_{\rm{u}} \rightarrow 750, N_{\rm{s}} \rightarrow 5000$    &     22.53&      0.26&      0.46&      2.93&      14.2\\ 
        B15.0\_1km\_0.5x\_1e2$^1$       & 1500 & -     & $\rho_\surf \rightarrow 10^2 \: \rm{g}$                      &     22.50&      0.25&      0.46&      2.94&      14.2\\ 
        B15.0\_1km\_0.5x\_2.0$^2$       & -    & -     & $\rho(t=0),P(t=0) \rightarrow \: \propto r^{-2}$             &     22.50&      0.25&      0.46&      2.93&      14.2\\ 
        B15.0\_1km\_0.5x\_11km$^3$      & -    & -     & $R_{\rm{in}} \rightarrow 11 \: \rm{km}$                      &     22.34&      0.29&      0.46&      3.00&      14.5\\ 
        B15.0\_1km\_0.5x\_10000km$^4$   & -    & -     & $R_{\rm{out}} \rightarrow 10000 \: \rm{km}$                  &     22.50&      0.26&      0.46&      2.93&      14.2\\ 
        B15.0\_1km\_0.5x\_floor$^5$     & -    & -     &$\rho_{\rm{floor}}\rightarrow 2 \times 10^{-22} \rho_{\rm ND}$&     22.50&      0.25&      0.46&      2.93&      14.2\\ 
        B15.0\_1km\_0.5x\_1.43$^6$      & -    & -     & $\Gamma \rightarrow \Gamma_* = 1.43$                         &     22.31&      0.24&      0.46&      2.30&      12.5\\ 
        B15.0\_1km\_0.5x\_rk3$^7$       & -    & -     & time stepping $\rightarrow$ 3rd-order Runge-Kutta            &     22.50&      0.25&      0.46&      2.93&      14.2\\ 
        B15.0\_1km\_0.5x\_two-sh$^8$    & -    & -     & Riemann solver $\rightarrow$ two-shock HLLC                  &     22.98&      0.31&      0.46&      2.91&      14.1\\ 
        B15.0\_1km\_0.5x\_par$^9$       & -    & -     & reconstruction $\rightarrow$ parabolic                       &     23.25&      0.47&      0.46&      3.27&      15.1\\ 
        B15.0\_1km\_two-sh              & -    & 10000 & solver + resolution                                          &     22.98&      0.31&      0.46&      2.92&      14.1\\ 
        B15.0\_1km\_2x\_par             & 3000 & -     & reconstruction + resolution                                  &     23.12&      0.50&      0.46&      3.32&      15.3\\
        \hline 
        \hline 
    \end{tabular}
\end{table*}

\subsection{Fiducial Model}\label{sec:fiducial}
\begin{figure*}
	\includegraphics[width=\textwidth]{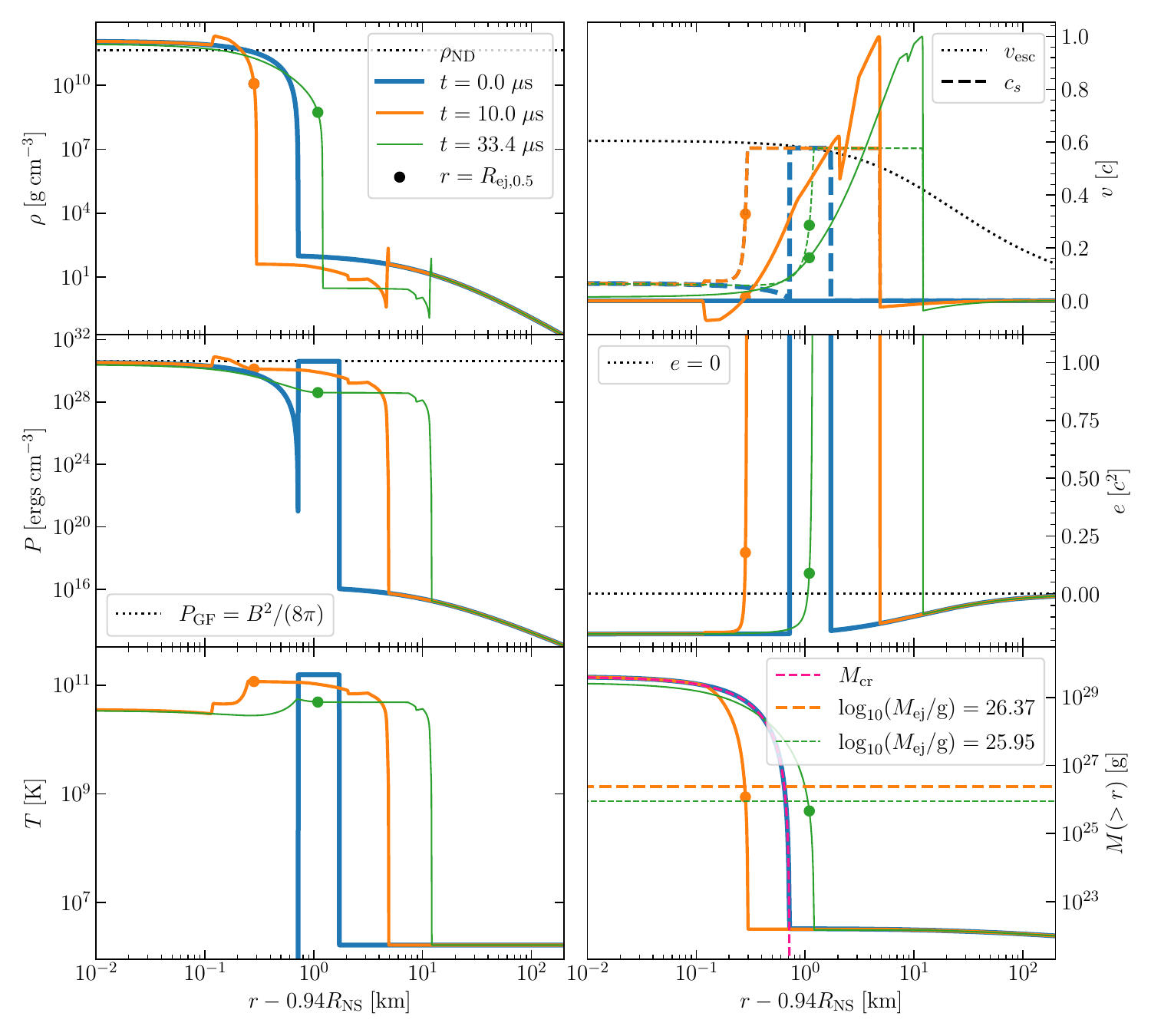}
    \caption{Radial profiles for the Fiducial model of quantities relevant to the ejecta dynamics, shown at three different early-time snapshots in time relative to the initialization of the high-pressure shell as marked.  These quantities include the density $\rho$, pressure $P$, temperature $T$, velocity $v$ (in comparison to the local escape velocity $v_{\rm esc}$ and sound speed $c_{\rm s}$), total energy density $e$ (equation \ref{eq:en_dens}), and the mass coordinate $M(>r)$ (equation \ref{eq:M_num}); the latter matches the analytic coordinate $M_{\rm cr}$ (equation \ref{eq:M_cr}) almost perfectly.  The radius exterior to which half the unbound ejecta mass (equation \ref{eq:R_ej,0.5}) is located are shown on each profile by solid dots.  On top of the cumulative mass profiles $M(>r)$ we show the original NS crust profile and two estimates of the total unbound ejecta mass $M_{\rm ej}$ (equation \ref{eq:M_ej}).}
    \label{fig:16.0-1.0-shrt-dyn}
\end{figure*}

\begin{figure*}
	\includegraphics[width=\textwidth]{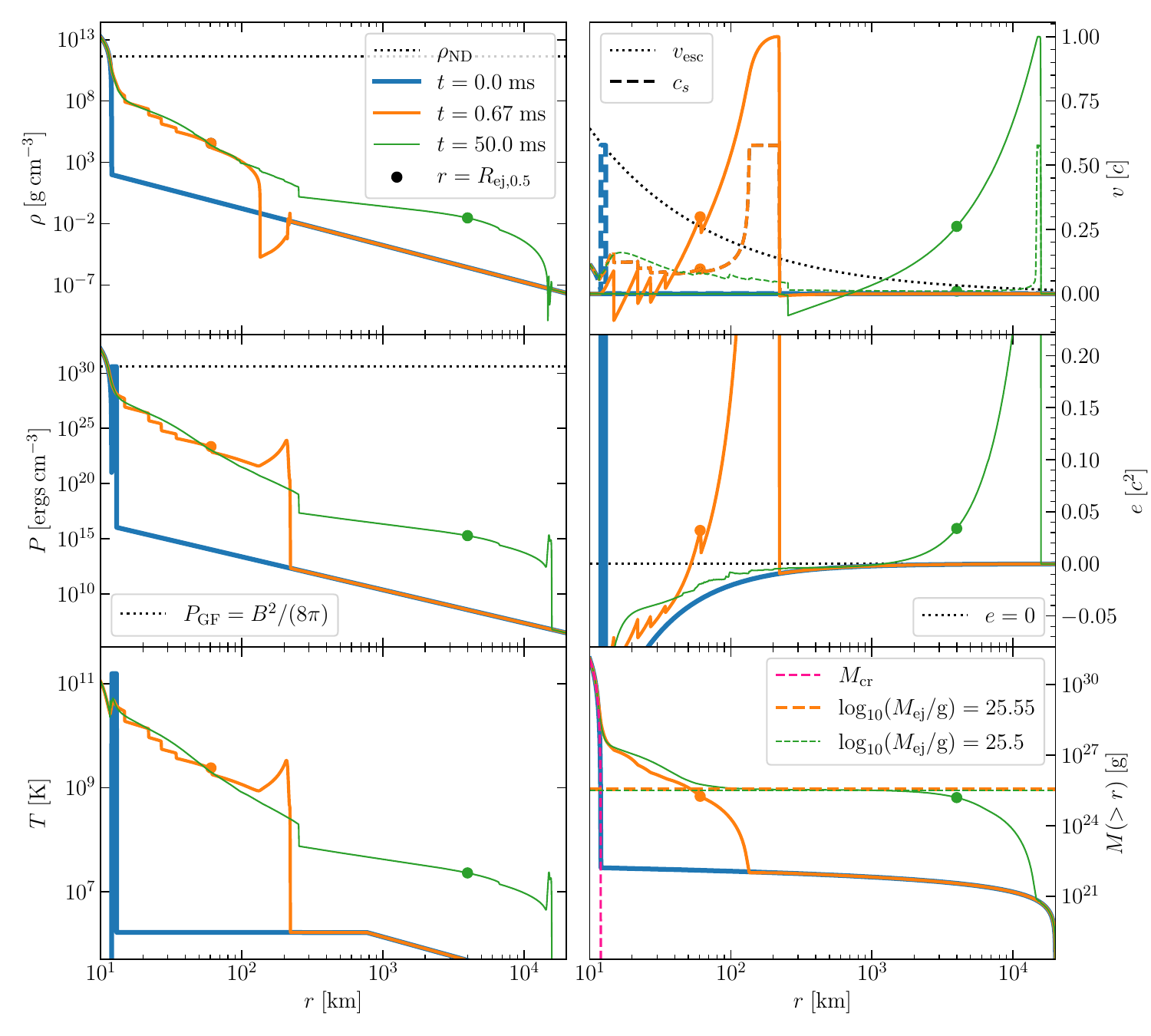}
    \caption{Radial profiles for the Fiducial model of quantities relevant to the ejecta dynamics, but now shown at three later snapshots on a zoomed-out radial grid.  The meaning of all quantities is the same as in Fig.~\ref{fig:16.0-1.0-shrt-dyn}.}
    \label{fig:16.0-1.0-ls-dyn}
\end{figure*}

\begin{figure*}
	\includegraphics[width=\textwidth]{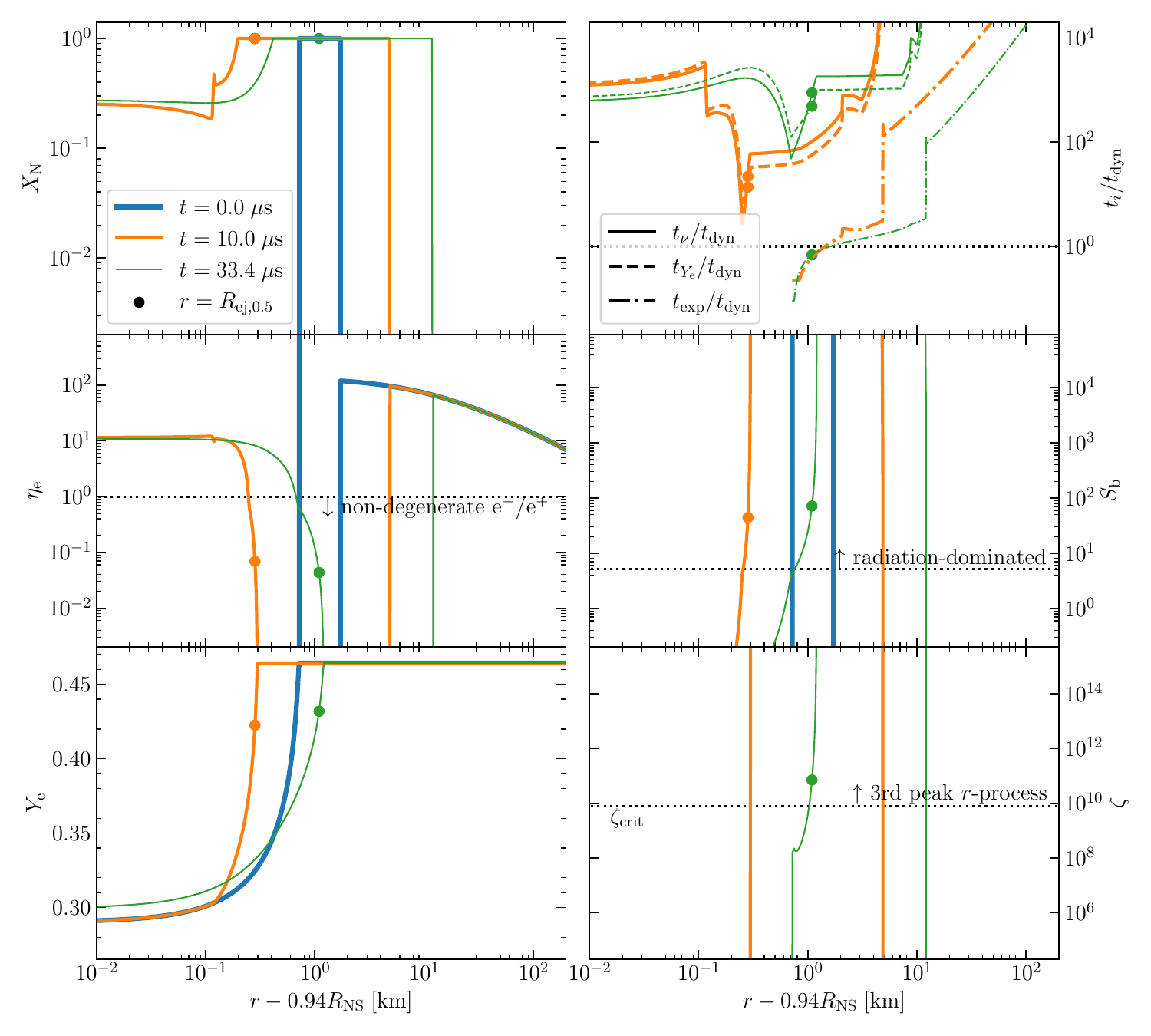}
    \caption{Radial profiles for the Fiducial model of quantities relevant to nucleosynthesis and neutrino emission, shown at the same three early-time snapshots shown in Fig.~\ref{fig:16.0-1.0-shrt-dyn}.  These quantities include: the free nucleon mass fraction $X_{\rm N}$ (equation \ref{eq:X_N}); degeneracy parameter $\eta_{\rm e} = \mu_e/(kT)$ (equation \ref{eq:eta_e}); electron fraction $Y_{\rm e}$; the ratios of different key timescales $t_i / t_{\rm dyn}$, where $t_i = t_\nu, t_{Y_{\rm e}}$, or $t_{\rm exp}$ (equations \ref{eq:t_dyn},\ref{eq:t_nu},\ref{eq:t_Y_e},\ref{eq:t_exp}, respectively); entropy per baryon $S_{\rm b}$ (equation \ref{eq:S_b}); and the $r$-process figure-of-merit parameter $\zeta$ (equation \ref{eq:zeta}).}
    \label{fig:16.0-1.0-shrt-nn}
\end{figure*}

\begin{figure*}
	\includegraphics[width=\textwidth]{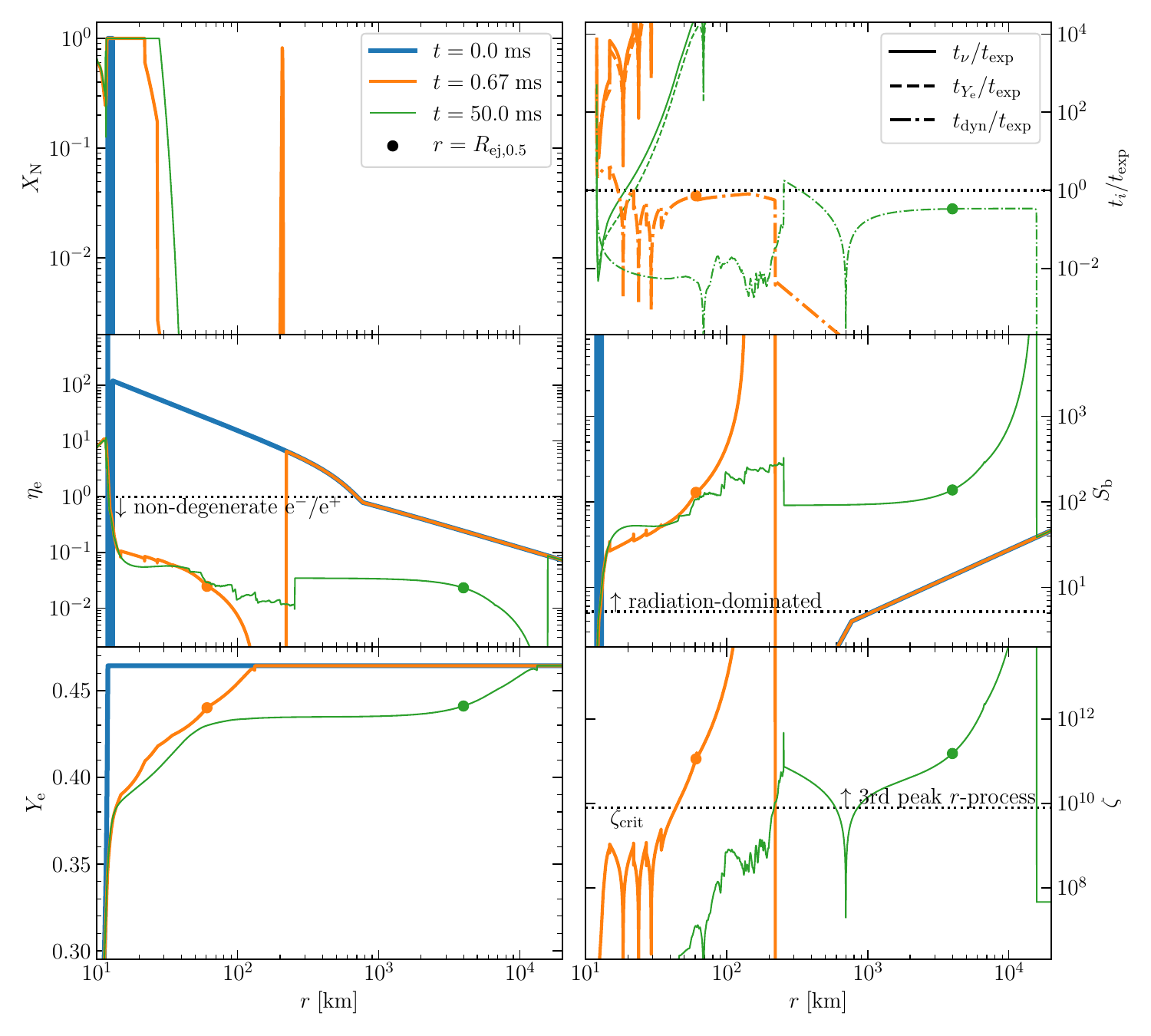}
    \caption{Radial profiles for the Fiducial model of quantities relevant to nucleosynthesis and neutrino emission, shown at three different late-time snapshots (same as in Fig.~\ref{fig:16.0-1.0-ls-dyn}) on a zoomed-out radial grid.  The meaning of all quantities is the same as in Fig.~\ref{fig:16.0-1.0-shrt-nn}.}
    \label{fig:16.0-1.0-ls-nn}
\end{figure*}

\begin{figure*}
	\includegraphics[width=\textwidth]{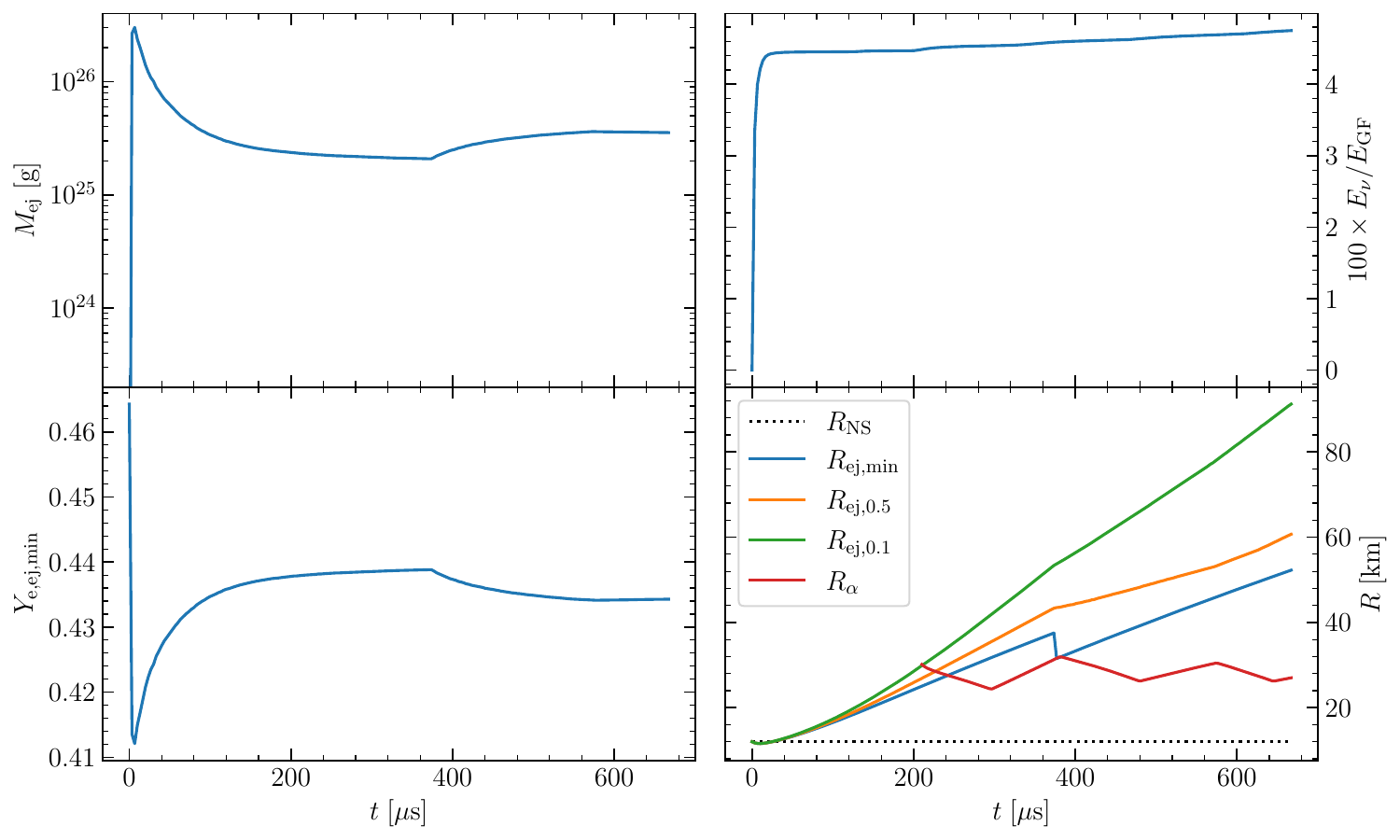}
    \caption{Time evolution of several quantities for the Fiducial model, including: unbound ejecta mass $M_{\rm ej}$ (defined as all layers with $e > 0$, equation \ref{eq:M_ej}); minimum electron fraction of the unbound ejecta $Y_{\rm e,ej,min}$; ratio of the total energy radiated in neutrinos $E_{\nu}$ (based on post-processing, equation \ref{eq:E_nu}) to the initial energy of the high-pressure sure $E_{\rm GF}$ (equation \ref{eq:E_GF}); and various critical radii defining the unbound ejecta layers, $R_{\rm ej,min}$ (equation \ref{eq:R_ej,min}), $R_{\rm ej,0.5}$ (equation \ref{eq:R_ej,0.5}), $R_{\rm ej,0.1}$, and an estimate of where $\alpha$-particle formation occurs $R_{\alpha}$ (equation \ref{eq:R_alpha}).  Here, $R_{\rm{ej},0.1}$ is defined in analogy to $R_{\rm{ej},0.5}$, as the radius exterior to which 10\% of the unbound ejecta mass resides.}
    \label{fig:16.0-1.0-shrt-time}
\end{figure*}

\begin{figure*}
	\includegraphics[width=0.96\textwidth]{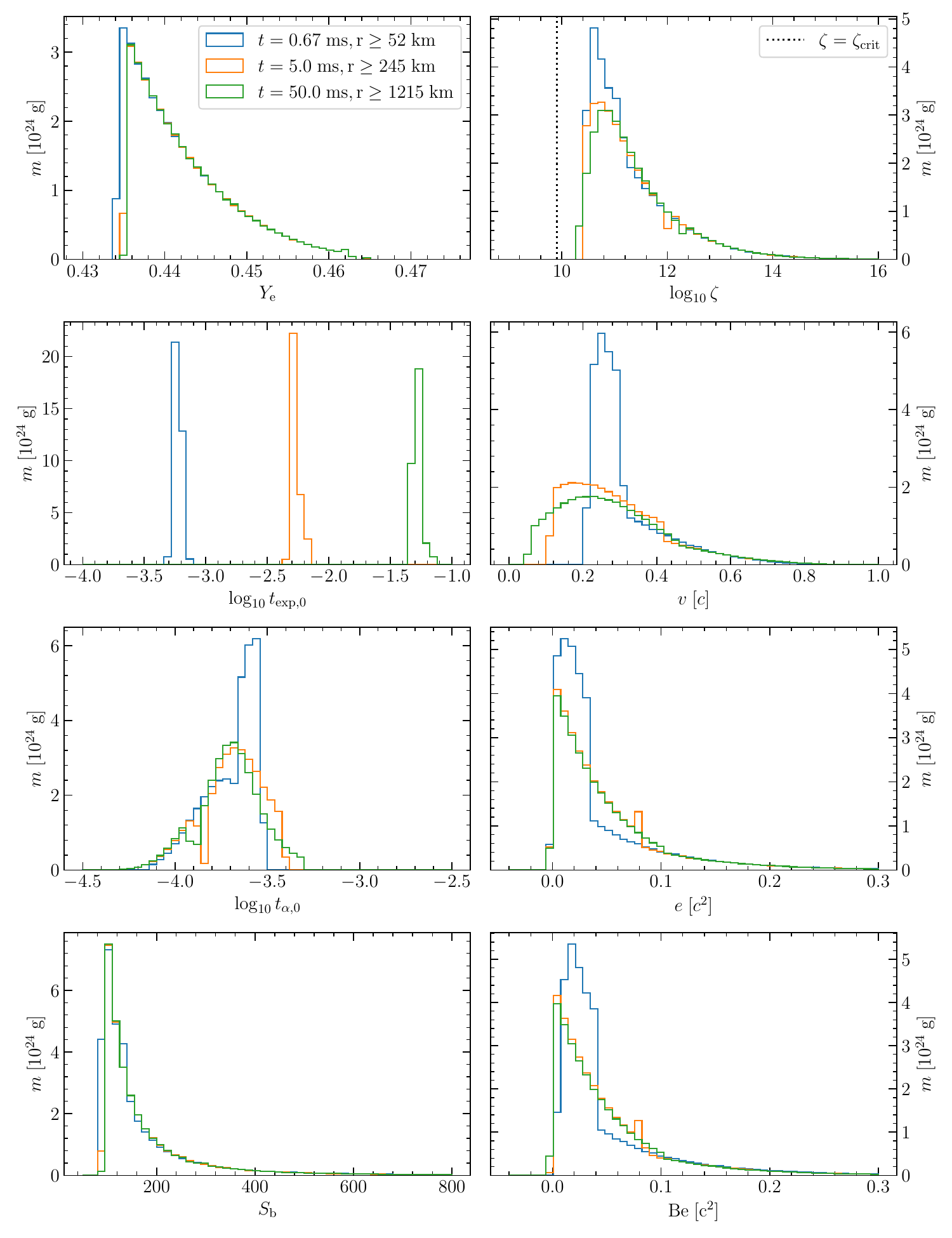}
    \caption{Mass-weighted distributions of various properties of the unbound ejecta (i.e., only those layers satisfying $e > 0$), at three different snaphots as labeled.  The quantities shown include: electron fraction $Y_{\rm e}$, radial expansion timescales $t_{\rm{exp},0}, t_{\alpha,0}$ (equations \ref{eq:t_exp},\ref{eq:t_exp,alpha}), entropy per baryon $S_{\rm b}$ (equation \ref{eq:S_b}), $r$-process figure-of-merit parameter $\zeta$ (equation \ref{eq:zeta}), velocity $v$, energy density $e$ (equation \ref{eq:en_dens}), and Bernoulli parameter $\text{Be} = e + P/\rho$.}
    \label{fig:16.0-1.0-ls-hist}
\end{figure*}

Figures~\ref{fig:16.0-1.0-shrt-dyn} and \ref{fig:16.0-1.0-shrt-nn} show early-time snapshots zoomed-in on the NS surface, of the radial profiles of several quantities relevant to the ejecta dynamics, nucleosynthesis, and neutrino emission, while Fig.~\ref{fig:16.0-1.0-shrt-time} shows the time evolution of several ejecta properties during this phase.  By ``early-times'' we mean shortly after the high-pressure shell has been initialized, when the NS material is being pushed inwards and shocked, then released due to decrease of pressure communicated from the outside by the rarefaction wave, and, subsequently, undergoes outwards re-expansion as the high thermal energy is transformed back into kinetic energy (analogy to an uncoiling spring).  This process takes $\rm{few} \times 10 \: \mu\rm{s}$ in our Fiducial simulation.  Later-time snapshots of the same quantities on a large radial grid, are shown in Figs.~\ref{fig:16.0-1.0-ls-dyn} and \ref{fig:16.0-1.0-ls-nn}, illustrating the expansion of the unbound ejecta away from the NS.  Mass-weighted distributions of several key properties of the unbound ejecta are shown in Fig.~\ref{fig:16.0-1.0-ls-hist}.

The shell-pressure of the Fiducial model $P_{\rm{GF}} = 4 \times 10^{30} \: \rm{ergs} \: \rm{cm}^{-3}$ corresponds to a uniform temperature $T_{\rm GF} \approx 2.0 \times 10^{11} \: \rm{K}$ in the radiation-dominated medium.  The second snapshot in Fig.~\ref{fig:16.0-1.0-shrt-dyn} ($t = 10.0 \: \mu \rm{s}$; orange lines) corresponds to the point of the greatest compression of the NS crust.  The density profile at this point is compressed inwards to around $0.4$ km below the initial NS radius, with $c_{\rm s} = c / \sqrt{3} \approx 0.58 c$ in the shocked region.  Considering that $10 \: \mu\rm{s} \cdot 0.58c \approx 1.7 \: \rm{km}$, the rarefaction wave launched inwards from the outer edge of the high-pressure region has by this time caught up with the shock being driven into the NS crust, reducing the pressure sufficiently to allow material to reverse its radial direction and to start rebounding outwards.  

The third snapshot in Fig.~\ref{fig:16.0-1.0-shrt-dyn} at $t = 33.4 \: \mu\rm{s}$ shows the expansion of the previously shocked material, as its thermal energy is converted back into the kinetic energy of outwards motion (the uncoiling of the spring; thin green lines).  This behavior is perhaps most easily visible in the density profile, which now extends hundreds of meters beyond the initial NS surface, as does the half-ejecta mass radius $R_{\rm{ej},0.5}$ (equation~\ref{eq:R_ej,0.5}).  The positive outwards velocity $v > 0$ of the shocked material is also apparent.  The fact that $e < 0$ at small $r < R_\ns$ by this point shows that no more unbound ejecta originating from the NS is expected.  Indeed, the ejecta shell is on the verge of being sonically disconnected from the NS surface (see also Fig.~\ref{fig:16.0-1.0-ls-dyn}).

The second snapshot in the late-time profiles shown in Fig.~\ref{fig:16.0-1.0-ls-dyn} corresponds to $t = 0.67 \: \rm{ms}$ (orange lines).  As can be seen most clearly in the density profile, the ultra-relativistic shock driven into the region outside the high-pressure shell is followed by a low-density cavity region and, subsequently, the shocked NS crust ejecta.  The latter has become sonically disconnected from the NS star already at this point, with the total ejecta mass $M_{\rm ej}$ nearly saturated at its final value $\approx 10^{25.5} \: \rm{g}$ (see also Table~\ref{tab:results}).  By the third snapshot at $t = 50 \: \rm{ms}$, the velocity profile decreases moving to smaller radii in the ejecta, before passing through zero becoming negative at $r < \rm{few} \times 100 \: \rm{km}$.  This position marks the location of the now-weakened shock, separating the subsonic shocked NS crust still tightly bound to the NS from the supersonically-expanding largely unbound ejecta layers.  The shock is also clearly visible in the density, pressure, and temperature profiles. 

Figs.~\ref{fig:16.0-1.0-shrt-nn} and \ref{fig:16.0-1.0-ls-nn} show three radial profile snapshots, this time of quantities relevant to the ejecta thermodynamics, nucleosynthesis and neutrino emission.  The early-time snapshots in Fig.~\ref{fig:16.0-1.0-shrt-nn} follow those in Fig.~\ref{fig:16.0-1.0-shrt-dyn}.  We first note that all of the shock-heated material (the two later snapshots; orange and green lines) that becomes unbound is dissociated into free nucleons ($X_{\rm N} = 1$).  These layers are also non-degenerate ($\eta_{\rm e} \ll 1$ at  $r \sim R_{\rm{ej},0.5}$), justifying our neglect of electron/positron degeneracy effects on the EOS and weak interaction rates.  We further see that while the shock reaches down to layers in the NS crust where $Y_{\rm e} \approx 0.30$, the unbound ejecta has a higher electron fraction $Y_{\rm e} \approx 0.43$ because it originates further out.  The hierarchy of timescales $t_\nu / t_{\rm dyn} \sim t_{Y_{\rm e}} / t_{\rm dyn} \sim 10^2$ at around $R_{\rm{ej},0.5}$ for the second snapshot (orange lines) and $t_\nu / t_{\rm dyn} \sim t_{Y_{\rm e}} / t_{\rm dyn} \sim 10^3$ near $R_{\rm{ej},0.5}$ for the third snapshot (green lines), have the implications that: (1) neutrino losses on the ejecta dynamics can be safely neglected (see also Fig.~\ref{fig:16.0-1.0-shrt-time}); (2) $Y_{\rm e}$ in the ejecta changes sufficiently slowly due to weak interactions that our assumption of using the original crustal $Y_{\rm e}$ is justified.  Finally, the fact that $t_{\rm exp} / t_{\rm dyn} \sim 1$ around $R_{\rm{ej},0.5}$ implies that our conclusions are insensitive to which definition of the expansion/dynamical timescale we use.  

The late-time snapshots shown in Fig.~\ref{fig:16.0-1.0-ls-nn} follow those in Fig.~\ref{fig:16.0-1.0-ls-dyn}.  Again, the unbound ejecta is seen to remain non-degenerate ($\eta_{\rm e} \ll 1$) throughout the entire simulation, and that its electron fraction saturates at $Y_{\rm e} \approx 0.44$ (see also Table~\ref{tab:results}).  By the third snapshot (green lines) we see that $t_\nu / t_{\rm exp}$ and $t_{Y_{\rm e}} / t_{\rm exp}$ become quite small $\sim 10^{-2}$ close to the NS surface, indicating that our neglect of neutrino losses and weak interactions are not self-consistent at small radii and late times.  However, because these inner bound layers are by now casually disconnected from the unbound ejecta shell, our predicted ejecta properties are not sensitive to changes in the late-time behavior that would result if neutrino cooling or weak interactions had been self-consistently included.  Finally, the entropy of the unbound ejecta is observed to achieve values $S_{\rm b} \sim \rm{few} \times 100$, corresponding to an $r$-process figure of merit parameter $\zeta \sim 10^{11}$ (see also Table~\ref{tab:results}); we return to the implications of this finding in Sec.~\ref{sec:rprocess}.

Fig.~\ref{fig:16.0-1.0-shrt-time} shows the time evolution of several key quantities over the first $\sim$0.7 ms of evolution.   The unbound ejecta mass $M_{\rm ej}$ increases rapidly initially, before changing course to decrease after $\sim 10 \: \mu\rm{s}$, and then increases again at $\sim 400 \: \mu \rm{s}$, ultimately saturating at a value $\simeq 3-4\times 10^{25} \: \rm{g}$ by $t \sim 600 \: \mu\rm{s}$.  The initial ``overshooting'' of $M_{\rm ej}$ occurs due to energy exchange between mass shells (i.e., layers which temporarily acquire $e > 0$ from the shock but later again become bound).  The later-time increase at 400\,$\mu$s is a result of a supersonic shock originating in the NS crust due to complex interplay between the original shock, the NS surface, and the inner reflecting boundary condition.  Subsequent weaker shocks originating in the NS crust manifest in Fig.~\ref{fig:16.0-1.0-ls-dyn} as discrete ``steps'' in various profiles at $\rm{few} \times 10$ km as shown with the orange line; however, these additional shocks do not significantly increase the unbound ejecta mass.  The behaviour of $Y_{\rm{e,ej,min}}$ follows that of $M_{\rm ej}$ as expected from the $Y_{\rm e,cr}(M_{\rm cr})$ profile of the NS crust.  While the total radiated neutrino energy $E_\nu$ increases sharply during the early shock phase ($\lesssim 10 \mu\rm{s}$), its final value reaches only a few percent of $E_{\rm GF}$ (achieved when $M_{\rm ej}$ saturates).  This again illustrates that neutrino losses are not important for the ejecta dynamics, at least for the Fiducial simulation (see Sec.~\ref{sec:many_sim}).  After $\sim 200 \: \mu\rm{s}$, the alpha-formation radius $R_\alpha$ in the ejecta saturates around a characteristic value $\approx 30$ km a couple NS radii above the surface.  The radii of the unbound ejecta $R_{\rm{ej,min}}, R_{\rm{ej},0.5}$, and $R_{\rm{ej},0.1}$ decrease slightly during the initial NS compression phase at $t \lesssim 10 \: \mu\rm{s}$,  (see Fig.~\ref{fig:16.0-1.0-shrt-dyn}), before increasing monotonically during the subsequent ejecta expansion and escape.  The radii $R_{\rm{ej,min}}, R_{\rm{ej},0.5}, R_{\rm{ej},0.1}$ intersect $R_\alpha$ at $t \approx 210 \: \mu\rm{s}, 227 \: \mu\rm{s}, 243 \: \mu\rm{s}$, respectively, causing the mass-weighted distribution of $\log_{10} t_{\alpha,0}$ to peak sharply at $-3.64$.  

Fig.~\ref{fig:16.0-1.0-ls-hist} shows mass-weighted distributions of quantities relevant to the ejecta dynamics and nucleosynthesis, at three late-time snapshots.  A comparison of the final two snapshots (orange and green lines, respectively) shows the mass-weighted distributions have largely saturated, i.e. the unbound ejecta is freely-expanding material with fixed properties.  This terminal electron fraction distribution samples values from the NS surface down to $Y_{\rm e} \approx 0.435$.  The expansion time distribution, $\log_{10} t_{\rm{exp},0}$, is narrowly concentrated at the simulation duration $t$, i.e. the time since the onset of the GF.  This is consistent with our assumption of free expansion when calculating $t_{\alpha,0}$, because for $v = \rm{const.}$ we have $t_{\rm{exp}} = \rm{const.}$ (equation \ref{eq:t_exp}; see also Fig.~\ref{fig:16.0-1.0-ls-nn}).  The unbound ejecta layers obey $\zeta > \zeta_{\rm crit}$, thereby satisfying the conditions for third-peak $r$-process nucleosynthesis (Sec.~\ref{sec:rprocess}).  The similarity between the distributions of $e$ and $\rm{Be}$ implies that a very similar result for $M_{\rm ej}$ is obtained whether one adopts an energy criterion ($e > 0$) or an alternative Bernoulli criterion ($\rm{Be} > 0$).  The qualitative shapes of the mass-weighted distributions are derived analytically in Appendix~\ref{app:analytic}. 

\subsection{Convergence Tests}\label{sec:convergence}
Table~\ref{tab:convergence} summarizes the results from several simulations performed as convergence checks, or other tests performed on one specific model, B15.0\_1km ($B = 10^{15}$ G, $P_{\rm GF} = 10^{28.6}$ ergs cm$^{-3}$, $\Delta R_{\rm GF} = 1 \: \rm{km}$).  The ejecta properties are found to be largely insensitive to: the adopted grid resolution, i.e. $N_{\rm u}$ and $N_{\rm s}$ (B15.0\_1km\_2x, B15.0\_1km\_0.5x, B15.0\_1km\_0.5x-0.5x); the NS surface density $\rho_\surf$ (B15.0\_1km\_0.5x\_1e2); the slope of the initial density and pressure profiles (B15.0\_1km\_0.5x\_2.0); the locations of the inner boundary $R_{\rm in}$ (B15.0\_1km\_0.5x\_11km) and outer boundary $R_{\rm out}$ (B15.0\_1km\_0.5x\_10000km); the minimum initial density and pressure values $\rho_{\rm floor}$ and $P_{\rm floor}$ (B15.0\_1km\_0.5x\_floor); and the time resolution (B15.0\_1km\_0.5x\_rk3).  On the other hand, more noticeable changes to the results arise from changing the assumed adiabatic index $\Gamma$ (B15.0\_1km\_0.5x\_1.43), the Riemann solver (B15.0\_1km\_0.5x\_two-sh), or the adopted reconstruction scheme (B15.0\_1km\_0.5x\_par).

Model B15.0\_1km\_0.5x\_1.43 shows that increasing $\Gamma$ from $4/3$ to $\Gamma_*=1.43$ (corresponding to the polytropic index of the cold crust) results in a slightly lower ejecta mass $M_{\rm ej}$, leaves $v_{\rm ej}$ practically untouched, lowers $S_{\rm b,ej}$, and thus $\zeta_{\rm ej}$, significantly by a factor of about $4$ and $50$, respectively.  However, physically $\Gamma \simeq 4/3$ is satisfied to high accuracy  in the radiation-dominated shocked ejecta anywhere that $S_{\rm b} \gg 1$.  Indeed, we have checked that $\Gamma = 4/3$ is satisfied to better than 0.05 per cent by post-processing model B15.0\_1km.

Switching the Riemann solver from TVD Lax-Friedrichs to two-shock HLLC (B15.0\_1km\_0.5x\_two-sh) increases $M_{\rm ej}$ by a factor of $3$, increases $v_{\rm ej}$ by around $0.06 c$, but leaves $S_{\rm b,ej}$ and $\zeta_{\rm ej}$ almost unchanged.  On the other hand, models run using the two-shock HLLC solver are not more sensitive to resolution changes (B15.0\_1km\_two-sh) than for those using TVD Lax-Friedrichs.  Switching from the linear to parabolic reconstruction scheme (B15.0\_1km\_0.5x\_par) increases $M_{\rm ej}$ by a factor of around $5-6$, increases $v_{\rm ej}$ by around $0.22 c$, and increases $\zeta_{\rm ej}$ by around one order of magnitude.  Also, the change to parabolic reconstruction makes the model more sensitive to resolution changes (B15.0\_1km\_2x\_par).  The mass-weighted distributions of the key ejecta properties for the four different models (B15.0\_1km, B15.0\_1km\_two-sh, B15.0\_1km\_2x, B15.0\_1km\_2x\_par) in Fig.~\ref{fig:hist_comp} (Appendix~\ref{app:hist_comp}) show that changing the resolution (B15.0\_1km\_2x) or solver (B15.0\_1km\_two-sh) gives behavior qualitatively similar to the default setup (B15.0\_1km), but changing the reconstruction (B15.0\_1km\_2x\_par) results in some apparently aphysical results, such as a double-peaked velocity distribution, leading us to disfavor this scheme. 

In summary, the degree to which our numerical results for the ejecta properties are conserved can be conservatively estimated based on differences that result from different assumed solvers.  As such, we conclude that $M_{\rm ej}$ is reliable to a factor $\lesssim 3$, $v_{\rm ej}$ to $\lesssim 0.05 c$, $\log_{10}S_{\rm b,ej}$ to few hundredths, and $\log_{10} \zeta_{\rm ej}$ to few tenths in the case of the B15.0\_1km model.  As a final sanity check, we note that a test simulation performed without a high-pressure shell ($P_{\rm GF} = 0$) gives rise to zero ejecta, i.e. $M_{\rm{ej}} = 0$.  Since we have not set up the atmosphere in strict hydrostatic equilibrium with the adopted $\Gamma_*$, we do see a time-dependent hydrodynamical re-arrangement of the mass.  The simulation runs until the NS atmosphere develops a region of very low density where the outer power-law profile is attached to the steep density gradient of the NS atmosphere.  This happens after about $\sim 30 \: \rm{ms}$ of evolution and is to be expected, because we use a value of $\Gamma = 4/3$ that is not consistent in the deep NS atmosphere.

\subsection{Analytic Constraints}\label{sec:analytic}

We now provide analytic estimates of the unbound ejecta mass, $M_{\rm{ej}}$, following the physical picture outlined in Sec.~\ref{sec:overview}.  Consistent with the 1D nature of our simulations, we assume spherically symmetric outflows occuring from all sides of the neutron star surface (solid angle $\Delta \Omega = 4\pi$); however, in the more realistic case that the outflows are confined to only a fraction of the total surface (Fig.~\ref{fig:cartoon}), these may be considered as estimates of the isotropic mass-loss rate and rescaled accordingly (see Sec.~\ref{sec:caveats} for further discussion).

An absolute upper limit on $M_{\rm{ej}}$ arises from the \emph{global} criterion that the total energy of the GF exceed the minimal kinetic energy required to unbind the ejecta from the gravitational well of the NS.  Equating $E_{\rm GF}$ (equation \ref{eq:E_GF}) to $M_{\rm{ej}} v_{\rm{esc}}^2/2$, where $v_{\rm{esc}} = \sqrt{2 G M_\ns / R_\ns}$ is the escape speed, we obtain:
\begin{equation}\label{eq:M_ej,max,gl}
    \begin{aligned}
        M_{\rm{ej,max,gl}} &\approx \frac{4 \pi R_\ns^3}{G M_\ns} (3P_{\rm{GF}}) \Delta R_{\rm{GF}} \\
        &\approx 3.5 \times 10^{28}\:{\rm g}\:R_{\ns,12}^{3}M_{\ns,1.4}^{-1}P_{\rm{GF},30} \Delta R_{\rm{GF},1},
    \end{aligned}
\end{equation}
where $R_{\ns,12} = R_\ns / (12 \: \rm{km})$ and $M_{\ns,1.4} = M_\ns / (1.4 \: \Msun)$.  

This global criterion corresponds to the assumption of a homogeneous energy distribution, such that each mass shell possesses just enough energy to escape.  A more stringent, but more realistic {\it local} upper limit on the ejecta mass arises by considering only those layers of the shocked crust which separately achieve positive energy.  Considering that the NS crust starts effectively from rest, this criterion reads $\epsilon + \phi \gtrsim 0$ (equation \ref{eq:en_dens}).  Relating this condition on the post-shock density $\rho_{\rm sh}$ of the marginally-ejected layer, $\rho_{\rm{sh}} \approx 7 \rho_{\rm ej} \lesssim P_{\rm{GF}}/(\Gamma-1)/(G M_\ns / R_\ns)$, to its original pressure (equation \ref{eq:polytrope}) and associated mass-coordinate (equation \ref{eq:M_cr}) in the crust, we obtain the \emph{local} upper limit on the ejecta mass:
\begin{equation}\label{eq:M_ej,max,loc}
    \begin{aligned}
        M_{\rm{ej,max,loc}} &\approx \frac{4 \pi R_\ns^4}{G M_\ns} P_* \lp \frac{1}{7} \frac{P_{\rm{GF}}/(\Gamma-1)}{G M_\ns / R_\ns} \frac{1}{\rho_*}\rp^{\Gamma_*} \\
        &\approx 8.5 \times 10^{25} \: \rm{g} \: R_{\ns,12}^{5.43} M_{\ns,1.4}^{-2.43} P_{\rm{GF},30}^{1.43},
    \end{aligned}
\end{equation}
where we have used $\rho_{\rm sh} \approx 7\rho_{\rm ej}$, corresponding to the jump conditions for a strong shock for a radiation-dominated $\Gamma = 4/3$ plasma in the Newtonian limit (\citealt{thompson1986}; their Eq.~24).

Even $M_{\rm{ej,max,loc}}$ represents only an upper limit on the ejecta mass, because the shock does not have infinite time to drive into the crust before the driving external pressure is relieved by the outwards expansion of the high-pressure shell, which thus limits the amount of material shock-heated to the positive energy.  As the shock propagates to higher densities deeper inside the crust, it weakens and propagates slower, enabling the rarefaction wave launched by the outwards expansion of the shell to catch up to the shock, relieving the pressure and allowing material to escape to infinity.  For a thin initial high-pressure region ($\Delta R_{\rm{GF}}\ll R_{\ns}$) this occurs roughly when $v_{\rm{sh}} \lesssim c / \sqrt{3} \lp = c_{\rm{s}} \rp$, where $v_{\rm{sh}}$ is the shock velocity in the rest frame of the unshocked NS crust, see also Fig.~\ref{fig:cartoon}.  Substituting this condition into the relativistic shock jump condition (\citealt{blandford1976}; their Eq.~8),
\begin{equation}\label{eq:strong_shock}
    \frac{3 P_{\rm{sh}}}{2 \rho_{\rm cr} c^2} = \gamma^2_{\rm{s}},
\end{equation}
where $P_{\rm{sh}}$ is the post-shock pressure and $\gamma_{\rm{s}}$ the shock's Lorentz factor in the NS frame, yields $\rho_{\rm{ej}} \gtrsim P_{\rm{sh}} / c^2$ as the criterion for a layer to be ejected.  Finally, assuming an approximate equality of pressure across the contact discontinuity which separates the shocked crust from the high-pressure shell,
\begin{equation}\label{eq:P_sh}
    P_{\rm{sh}} = \alpha P_{\rm{GF}},
\end{equation}
where $\alpha \in \langle 0.1,1 \rangle$ is some constant of order unity, we obtain a {\it local} lower limit on the ejecta mass:
\begin{equation}\label{eq:M_ej,min,loc}
    \begin{aligned}
        M_{\rm{ej,min,loc}} &\approx \frac{4 \pi R_\ns^4}{G M_\ns} P_* \lp \frac{\alpha P_{\rm{GF}}}{c^2} \frac{1}{\rho_*}\rp^{\Gamma_*} \\
        &\approx  2.3 \times 10^{25} \: \rm{g} \: R_{\ns,12}^4 M_{\ns,1.4}^{-1} (\alpha P_{\rm{GF},30})^{1.43}.
    \end{aligned}
\end{equation}
Taking the ratio of equations (\ref{eq:M_ej,max,loc}) and (\ref{eq:M_ej,min,loc}) yields
\begin{equation}
    \frac{M_{\rm{ej,max,loc}}}{M_{\rm{ej,min,loc}}} = \lp \frac{c^2}{7 \alpha (\Gamma-1) G M_\ns / R_\ns}\rp^{\Gamma_*} \approx 10,
\end{equation}
for $R_{\ns,12} = 1$, $M_{\ns,1.4} = 1$, and $\alpha = 0.5$.  As we shall show in Sec.~\ref{sec:many_sim}, $M_{\rm{ej,min,loc}}$ assuming $\alpha = 0.5$ provides a good fit to the normalization and functional form of the ejecta masses $M_{\rm{ej}}(P_{\rm GF})$ measured from our suite of numerical simulations (equation \ref{eq:M_ej}) for $\Delta R_{\rm{GF}} = 2 \: \rm{km}$ (see Fig.~\ref{fig:M_ej}).  

One of the key quantities relevant to nucleosynthesis in the ejecta is the entropy it acquires from the shock (Sec.~\ref{sec:rprocess}).  The entropy of the shocked layer can be estimated as (equation~\ref{eq:S_b})
\begin{equation}
S_{\rm b,sh} \simeq 5.21\frac{T_{\rm sh,MeV}^{3}}{\rho_{\rm sh,8}} \simeq 627\frac{P_{\rm{GF},30}^{3/4}}{\rho_{\rm cr,8}},
\label{eq:S_sh_analytic}
\end{equation}
where $\rho_{\rm sh,8} = 7\rho_{\rm cr,8}$ and $T_{\rm sh} \simeq (12P_{\rm GF}/11a)^{1/4}$ are the density and temperature of the post-shock gas, respectively.  

\begin{figure}
	\includegraphics[width=\columnwidth]{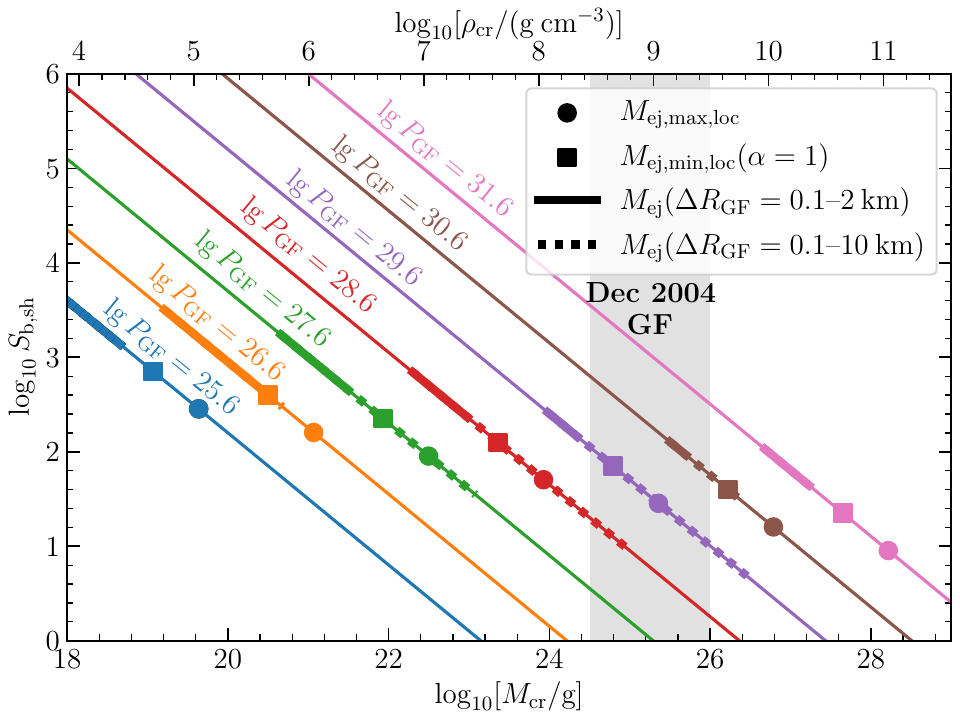}
    \caption{Analytic estimate of entropy obtained behind the shock driven into the NS crust, $S_{\rm b,sh}$ (equation \ref{eq:S_sh_analytic}), as a function of the density, or, equivalently, mass-depth $M_{\rm cr}$, of the crust, for different values of the externally applied shell-pressure $P_{\rm GF}$ (different colored lines, as marked).  Shown for comparison with symbols are analytic estimates for the total unbound ejecta mass (equations \ref{eq:M_ej,max,loc}, \ref{eq:M_ej,min,loc}) as well as the ejecta masses $M_{\rm ej}$ measured from our simulations for a range of initial shell-thicknesses, $\Delta R_{\rm GF}$.  For models capable of reproducing the baryon ejecta mass inferred from the Dec. 2004 GF of SGR1806-20 (vertical gray-shaded region), we should expect ejecta entropies $S_{\rm b,sh} \gtrsim 100$ in the range necessary for $r$-process nucleosynthesis (Sec.~\ref{sec:rprocess}).}
    \label{fig:S_b}
\end{figure}

In Fig.~\ref{fig:S_b} we show $S_{\rm b,sh}$ as a function of $M_{\rm cr}$ obtained from $\rho_{\rm cr}(M_{\rm{cr}})$ (Fig.~\ref{fig:P-Y_e-H_rho}) for different values of $P_{\rm GF}.$  We see that for sufficiently powerful flares ($P_{\rm GF} > 10^{30}$ ergs cm$^{-3}$) to unbind $M_{\rm cr} \gtrsim 10^{24.5}$ g according to even our conservative local criteria, entropies of $S_{\rm b,sh} \gtrsim 100$ can be achieved in these ejecta layers.  Such entropy values can be sufficient for third-peak $r$-process nucleosynthesis (Sec.~\ref{sec:rprocess}). 

\subsection{Parameter Study}\label{sec:many_sim}
In this section we compare the analytic constraints derived in Sec.~\ref{sec:analytic} with simulation results shown in Table~\ref{tab:results}. 
\begin{figure*}
	\includegraphics[width=\textwidth]{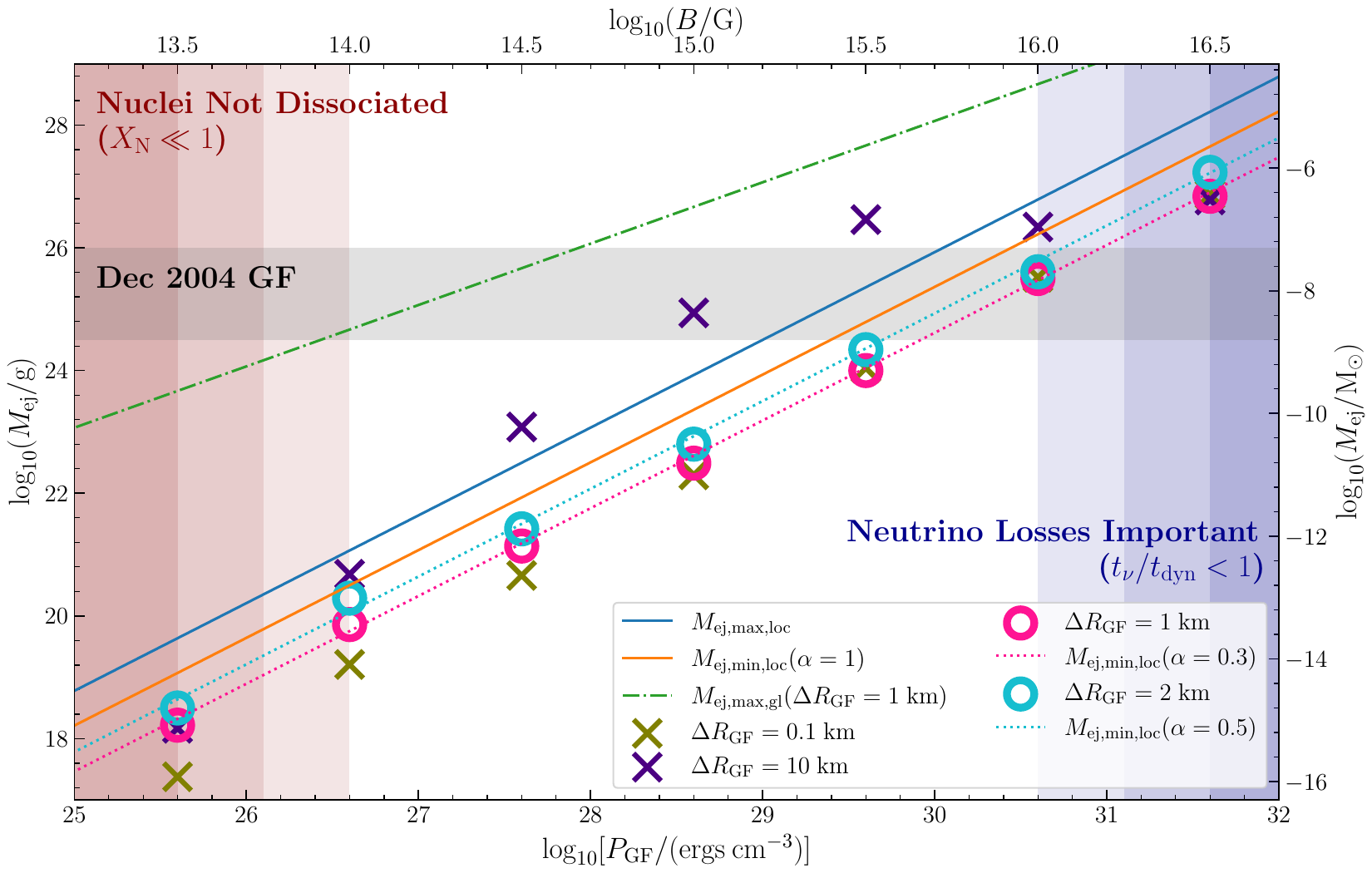}
    \caption{Unbound ejecta mass $M_{\rm ej}$ from our suite of simulations as a function of the initial shell-pressure $P_{\rm GF}$ (equivalently, strength of the dissipated magnetic field $B$; top horizontal axis) for different initial shell thicknesses $\Delta R_{\rm GF}$ as indicated.  Shown for comparison are analytic estimates bounding the ejecta mass (see Sec.~\ref{sec:analytic}).  The range of ejecta masses inferred from the radio afterglow of the Dec.~2004 GF of SGR 1806-20 \citep{granot2006} is shown with gray shading (Sec.~\ref{sec:afterglow}).}
    \label{fig:M_ej}
\end{figure*}

In Fig.~\ref{fig:M_ej} we show our results for the total ejecta mass $M_{\rm ej}$ from four sets of simulations with $\Delta R_{\rm{GF},1} \in \{ 0.1, 1, 2, 10\}$ and other model parameters according to Table~\ref{tab:results}.  For comparison we show analytic estimates of the maximal mass based  the \emph{global} energy criterion $M_{\rm ej,max,gl}$ (equation \ref{eq:M_ej,max,gl}) and the \emph{local} energy criterion $M_{\rm ej,max,loc}$ (equation \ref{eq:M_ej,max,loc}) together with the minimal mass $M_{\rm ej,min,loc}$ (equation \ref{eq:M_ej,min,loc}).  Red and blue shaded regions correspond to regions of parameter space where the shock energy is too weak to dissociate the ejecta into free nucleons (i.e. $X_{\rm N} \ll 1$; equation \ref{eq:X_N}), or too strong for neutrino cooling to negligibly influence the ejecta dynamics (i.e. $t_\nu / t_{\rm dyn} < 1$; equations \ref{eq:t_nu},\ref{eq:t_dyn}), respectively.  The quantities $X_{\rm N}$ and $t_\nu / t_{\rm dyn}$ are measured for the models with $\Delta R_{\rm Gf} = 1 \: \rm{km}$ at times close to the point of the greatest compression of the NS surface ($t = 10.0 \: \mu\rm{s}$) and at the half-mass radius $r = R_{\rm{ej},0.5}$.  The measured values $X_{\rm N} (\text{B13.5\_1km}) \sim 4 \times 10^{-2}$, $X_{\rm N} (\text{B14.0\_1km}) \approx 1$, $t_\nu / t_{\rm dyn}(\text{B16.0\_1km}) \sim 50$, and $t_\nu / t_{\rm dyn}(\text{B16.5\_1km}) \sim 2 \times 10^{-1}$ inform the painted regions in Fig.~\ref{fig:M_ej}.  Thus, for $B \lesssim 10^{13.5} \: \rm{G}$ $(P_{\rm GF} \lesssim 10^{25.5}$ ergs cm$^{-3}$) and $B \gtrsim 10^{16.5}$ $(P_{\rm GF} \gtrsim 10^{31.5}$ ergs cm$^{-3}$), our physical assumptions break down and the accuracy of the simulation results are questionable.

As discussed in Sec.~\ref{sec:analytic}, the ejecta mass $M_{\rm{ej,max,gl}} \propto P_{\rm{GF}}$ based on the \emph{global} energy criterion (equation \ref{eq:M_ej,max,gl}) for $\Delta R_{\rm{GF}} = 1 \: \rm{km}$ is much less restrictive than the \emph{local} energy constraint $M_{\rm{ej,max,loc}} \propto P_{\rm{GF}}^{\Gamma_*}$ (equation \ref{eq:M_ej,max,loc}).  The dependence of our numerically calculated $M_{\rm{ej}}$ (equation \ref{eq:M_ej}) on $P_{\rm GF}$ more or less follows the trend predicted by the minimum ejecta mass $M_{\rm{ej,min,loc}} \propto P_{\rm{GF}}^{\Gamma_*}$ estimate derived from the \emph{local} energy criterion (equation \ref{eq:M_ej,min,loc}).  We return to the implications of our results for the December 2004 GF of SGR 1806-20 in Sec.~\ref{sec:afterglow}.

\section{Summary and discussion}\label{sec:discussion}

Using 1D RHD simulations and analytic estimates, we have investigated the hypothesis that the sudden dissipation of magnetic energy above the NS surface in a magnetar GF drives a relativistic shockwave into the NS crust, ultimately giving rise to the ejection of neutron-rich material (Fig.~\ref{fig:cartoon}).  Our results for the system evolution and ejecta properties are summarized in Figs.~\ref{fig:16.0-1.0-shrt-time}-\ref{fig:M_ej}.  The videos produced from the Fiducial model are publicly available in \href{https://zenodo.org/records/10593900}{Zenodo}, on \href{https://www.youtube.com/playlist?list=PLtfS1xXqUmtC2A48QhRvGXxbeMK1vkNnC}{YouTube}, and alongside the paper as supplementary material on the journal website.  We now describe the implications of our results for GF radio afterglows, $r$-process nucleosynthesis, and FRBs.  We conclude by highlighting some caveats and limitations of our numerical calculations and considering the prospects for future directions.

\subsection{Radio Afterglows of Magnetar GFs}\label{sec:afterglow}
Assuming a spherical or nearly spherical outflow covering a large fraction of the NS surface, the ejecta masses $M_{\rm{ej}} \gtrsim 10^{24.5} \: \rm{g}$ and kinetic energies we find are consistent with those inferred from the radio afterglow of the Dec.~2004 SGR 1806-20 GF \citep{granot2006,gelfand2005} for an assumed initial shell pressure $P_{\rm{GF}} \gtrsim 10^{30} \: \rm{ergs} \: \rm{cm}^{-3}$ (Fig.~\ref{fig:M_ej}), corresponding to a dissipated magnetic field of strength $B \gtrsim 10^{15.5} \: \rm{G}$ (equation \ref{eq:P_shock}).  The latter is indeed similar to the dipole field strength $B_{\rm dip} \approx 2\times 10^{15}$ G measured from the spin-down rate of SGR 1806-20 \citep{Olausen&Kaspi14}. 

However, matching the thermal energy contained in the high-pressure shell (equation \ref{eq:E_GF}) to the $\gtrsim 2-4\times 10^{46}$\,ergs of gamma-ray emission observed from the 2004 GF \citep{palmer2005,hurley2005} would appear to require a very thin shell $\lesssim 0.01$ km, possibly pointing to some tension with the model or the observations.  For example, the true energy budget of the 2004 GF might be considerably higher than implied by the observed (isotropic) gamma-ray luminosity.  We further discuss this tension and another possible resolution involving an aspherical outflow in Sec.~\ref{sec:caveats}.

\subsection{\textit{r}-Process Nucleosynthesis}\label{sec:rprocess}
Our finding that magnetar GFs can give rise to the ejection of significant quantities of NS crust material has implications for the synthesis of $r$-process elements.  Expanding initially cold NS matter was suggested as an $r$-process site by \citet{lattimer1974,lattimer1977}.  Here the ejecta is far from cold: the GF-driven shock heats up the NS crustal material sufficiently to dissociate the heavy nuclei into free nucleons ($X_{\rm N} \simeq 1$; e.g., Fig.~\ref{fig:16.0-1.0-shrt-nn}) for $P_{\rm{GF}} \gtrsim 10^{25.5} \: \rm{ergs} \: \rm{cm}^{-3}$ ($B \gtrsim 10^{13.5} \: \rm{G}$; see Fig.~\ref{fig:M_ej}).  Nevertheless, the temperatures achieved are not sufficiently high for weak interactions, particularly positron captures on neutrons, to appreciably increase the electron fraction from the neutron-rich composition $Y_{e} \sim 0.4-0.5$ of the pre-flare crust.  We furthermore find that the entropy attained by the ejecta is sufficiently high (Fig.~\ref{fig:S_b}), and its expansion rate through the seed-nuclei formation region sufficiently fast  ($\zeta \gtrsim \zeta_{\rm crit}$; equation \ref{eq:zeta}) for an alpha-rich freeze-out \citep{hoffman1997}, allowing neutron captures to proceed up to and even beyond the third $r$-process peak (e.g., Fig.~\ref{fig:16.0-1.0-ls-nn}).  Our results therefore support magnetar GFs as a new astrophysical site for the $r$-process.  

How important are magnetar GF to Galactic chemical evolution as a whole?  The average production rates in the Milky Way for all $r$-process elements, as well as just those above the second ($A > 130$) and third ($A > 195$) $r$-process peaks, are roughly given by $\dot{M}_{\rm{r,all}} \approx 1.9 \times 10^{-6} \: \Msun \: \rm{yr}^{-1}$, $\dot{M}_{\text{r},A > 130} \approx 2.5 \times 10^{-7} \: \Msun \: \rm{yr}^{-1}$, and $\dot{M}_{\text{r},A > 195} \approx 5 \times 10^{-8} \: \Msun \: \rm{yr}^{-1}$, respectively \citep{rosswog2017}.  The rate of Galactic magnetar GFs is poorly constrained observationally, but if we assume that SGR 1806-20-like flares of energy $\gtrsim 10^{46}$ ergs occur at a rate of $\Gamma_{\rm GF} \sim 10^{-2}$\:yr$^{-1}$ (\citealt{beniamini2019}), then the corresponding GF $r$-process production rate is $\dot{M}_{\rm{r,GF}} \sim 10^{-9} \: \Msun \: \text{yr}^{-1} \: [\Gamma_{\rm GF} / (10^{-2}\: \rm{yr}^{-1})]$, assuming $M_{\rm ej} \sim 10^{26}$\,g of $r$-process ejecta per flare \citep{granot2006}.  While this corresponds to a negligible fraction of $\dot{M}_{\rm{r,all}}$, it represents a few percent of $\dot{M}_{\rm{r},A > 195}$, i.e. of those elements particularly difficult to synthesize in neutrino-driven core-collapse supernovae (e.g., \citealt{thompson2001}).  The yield estimated based on Galactic magnetars could furthermore represent a lower limit if a rare sub-population of magnetars (e.g., particularly young or highly-magnetized sources) are even more active than Galactic magnetars, as hinted by the existence of very active repeating FRB sources (see Sec.~\ref{sec:FRB} for further discussion).

Even a subdominant heavy $r$-process production which occurs promptly after star formation (as satisfied by GF given the young ages of Galactic magnetars; $\lesssim 10^{4}$ yr) could play an important role in the Galactic chemical evolution through enrichment of the most metal-poor stars in the Milky Way or its satellite dwarf galaxies (e.g., \citealt{Ji+16}).  Several studies have suggested that another channel (in addition to the established site of NS binary mergers) with short delay time relative to star-formation is needed to explain the [Eu/Fe]-[Fe/H] observations \citep[e.g.,][]{hotokezaka2018,Cote+19,Siegel+19,vanderswaelmen2023,lian2023}.  Channels that might satisfy this requirement include ``magnetorotational'' supernovae (e.g., \citealt{winteler2012,nishimura2015}; see however \citealt{moesta2018}), disk outflows from hyper-accreting black holes (e.g., \citealt{MacFadyen&Woosley99,Siegel+19}), and proto-magnetar neutrino-driven winds \citep{thompson2018}.  We suggest magnetar GFs be added to the list of potential prompt enrichment sites.

\subsection{Kilonova-like optical transients from magnetar GFs}\label{sec:kilonova}
As in the case of NS binary mergers, the ejection of $r$-process nuclei from magnetar GF flares could power a short-lived kilonova-like transient powered by radioactive decay (\citealt{Li&Paczynski98,Metzger+10}).  The transient peaks on the diffusion timescale,
\begin{equation}
t_{\rm pk} \approx \sqrt{\frac{M_{\rm ej}\kappa}{4\pi v_{\rm ej}c}} \simeq 300 \: {\rm s} \: \left(\frac{M_{\rm ej}}{10^{26}\:{\rm g}}\right)^{1/2}\left(\frac{v_{\rm ej}}{0.3c}\right)^{-1/2}\left(\frac{\kappa}{\rm 3\:cm^{2}\:g^{-1}}\right)^{1/2},
\end{equation}
where $\kappa$ is the opacity of $r$-process nuclei (e.g., \citealt{Tanaka+20}).  The peak luminosity can be roughly estimated from the radioactive heating rate of $r$-process nuclei on timescales $\sim t_{\rm pk}$ (\citealt{Metzger+10,Metzger19}),
\begin{equation}
L_{\rm pk} \approx 10^{39}\:{\rm ergs \:s^{-1}}\left(\frac{M_{\rm ej}}{10^{26}\:\rm g}\right)^{0.35} \left(\frac{v_{\rm ej}}{0.3c}\right)^{0.65} \left(\frac{\kappa}{\rm 3\:cm^{2}\:g^{-1}}\right)^{-0.65} .
\end{equation}
The effective temperature of the emission, $T_{\rm eff}\simeq(L_{\rm pk}/4\pi\sigma_{\rm SB}(v_{\rm ej}t_{\rm pk})^{2})^{1/4}\approx2\times10^{4}$ K, where $\sigma_{\rm SB}$ is the Stefan–Boltzmann constant, corresponds to optical/UV wavelength emission.  At an 15 kpc assumed distance of SGR 1806-20, $L_{\rm pk}$ corresponds to a peak apparent g-band AB magnitude $m_{\rm AB} \approx 7.$ Thus, we predict that substantial baryon ejection in Galactic GF would be accompanied by short $\lesssim$ minutes-long bright optical flare, a transient event we dub ``nova brevis'' (a brief/short nova).  Robotic telescopes with very large instantaneous fields of view covering much of the entire night sky (e.g., Evryscope; \citealt{Law+14}) appear best equipped to detect such bright, but very rare flares.  Alternatively, for extragalactic magnetar GFs, a rapidly slewing space-based optical/UV telescope, which triggers on the short gamma-ray spike of the GF, could potentially detect the nova brevis signal for events in the relatively nearby universe $\lesssim 10$ Mpc. Several UV satellite missions of this type, which can slew on a timescale of minutes with a sufficiently large field of view to encompass the gamma-ray error region, are currently planned, including ULTRASAT \citep{Sagiv+14}, QUVIK \citep{Werner+23}, or UVEX \citep{Kulkarni+21}.

\subsection{Fast Radio Burst Environments}\label{sec:FRB}
Even prior to the discovery of an FRB from a flaring Galactic magnetar \citep{Bochenek+20, CHIME+20}, extragalactic magnetars were considered the most promising models for the central engines of FRBs (e.g., \citealt{Lyubarsky14,Metzger+17,Beloborodov17}; see \citealt{lyubarsky2021} for a review).  Two of the best studied and most active repeating sources, FRB 121102 \citep{Chatterjee+17} and  FRB 190520B \citep{Niu+22}, are spatially coincident with luminous synchrotron point sources.  The sub-parsec sizes and high luminosities of these sources are consistent with them being young ($\sim$ decades-centuries old) magnetized nebulae filled with relativistic electrons, the latter accelerated at the termination shock of a quasi-steady trans-relativistic ``wind'' created by the accumulation of ejecta from magnetar flares \citep{Metzger+17,Beloborodov17,margalit2018}.  The extremely high and time-variable rotation measure of the bursts from FRB 121102 \citep{Michilli+18} can then be explained by the bursts passing through the same magnetized turbulent nebula responsible for the persistent synchrotron emission, but only provided that the flare ejecta has an electron-ion composition (a nebula of electron/positron pairs would impart no net rotation measure).  \citet{margalit2018} showed that all the basic properties of the FRB 121102 persistent source (size, flux, self-absorption constraints) and the large but decreasing rotation measure (RM) of the bursts, can be explained by a magnetar injecting a transrelativistic outflow with a velocity $v \sim 0.5 c$ and time-averaged mass flux $\dot{M} \sim 10^{19}-10^{21}$ g s$^{-1}$.  Given that the rarest, most powerful FRBs from FRB 121102 (perhaps analogs to magnetar GF) occur roughly once per day on average (e.g., \citealt{Nicholl+17}), this $\dot{M}$ corresponds to a per-flare baryon ejection of $\sim 10^{24}-10^{26}$ g, similar to that inferred for for SGR 1806-20 GF \citep{granot2006}.  This scenario predicts that repeating FRB sources may go ``dark'' for a period of hours to days after powerful bursts (those generated by GF-energy events) as a result of any subsequent FRBs undergoing absorption or induced Compton scattering by the baryon shell \citep{Metzger+19}.

Taken together, our results suggest that young active FRB sources like FRB 121102 may be generating a high mass-flux $\dot{M}$ of heavy $r$-process nuclei.  Over the $\Delta T \gtrsim 10$ yr active lifetime of FRB 121102, this would correspond to an $r$-process yield of $M_{\rm r} \sim \dot{M}\Delta T \sim 10^{-6}-10^{-4}M_{\odot}$, coincidentally similar to the potential $r$-process yield of the proto-neutron star wind phase (e.g., \citealt{thompson2001}).  However, the estimated birthrate of repeating FRB sources suggest that only a small fraction $\lesssim 1\%$ of all magnetars are as active as FRB 121102 (e.g., \citealt{Margalit+20}).  Within a scenario for $r$-process production from magnetar GF, a key open question is the relative contributions from the hyper-active magnetars responsible for repeating FRB sources and comparatively inactive magnetars like those in our Galaxy.  The possibility of a high rate of magnetar crust removal due to repeated GF also raises the question of whether the crustal composition will have time to maintain $\beta-$equilibrium (as assumed in our crust profile; Fig.~\ref{fig:P-Y_e-H_rho}), or whether the layers being excavated are even more neutron-rich than we have assumed.  Future work should explore the long-term evolution of the NS crust subject to rapid mass-loss, using nuclear reaction network calculations similar to those used to study the opposite process of mass {\it accretion} (e.g., \citealt{Schatz+99}).

\subsection{Simplifications, Caveats, and Future Work}\label{sec:caveats}
Let us now address some of the simplifications and idealizations employed in this work.  Firstly, we have neglected radiation losses; while a good approximation for photons in the highly optically-thick shock and outflow, neutrino losses can become relevant at high temperatures and densities such as those achieved in the shocked crust.  We have explored neutrino losses in post-processing, finding that they are only important for the flare dynamics for extremely high-pressures $P_{\rm{GF}} \gtrsim 10^{31.5} \: \rm{ergs} \: \rm{cm}^{-3}$ ($B \gtrsim 10^{16.5} \: \rm{G}$; blue shaded region in Fig.~\ref{fig:M_ej}).  

A more significant caveat is that we do not include magnetic fields in our simulations.  Strong magnetic fields (magnetization $\sigma_{\rm m} > 1$) qualitatively change the dynamics of the plasma as a result of flux freezing and the strong Lorentz force \citep{lyutikov2022,barkov2022}.  Although the energy source of magnetar flares is clearly the magnetic field, we are implicitly assuming that immediately following the flare, the efficiency of magnetic dissipation is high enough that the plasma covering at least a portion of the NS surface is comparatively weakly magnetized.  If the magnetic field in the crust remains sufficiently high that $\sigma_{\rm m} \gtrsim 1,$ then the transverse component of the field may suppress the formation of strong shocks, resulting in weaker heating of the plasma than we have assumed.  Future RMHD simulations are necessary to explore the effects of strong magnetic fields on our results.   

Our simulations are only 1D and hence implicitly assume spherical symmetry, despite the fact that the radio observations of the December 2004 GF indicate one-sided outflow on large scales with a 2:1 axis ratio \citep{taylor2005}.  In the physical situation, the high-pressure region, unconfined by the magnetic field, may cover only a small fraction of the NS surface $f_{\rm open} = \Delta \Omega/4\pi \ll 1$.  While for a fixed shell-pressure $P_{\rm GF}$, our 1D analytic estimates predict the total ejecta mass to scale as $M_{\rm ej} \propto f_{\rm open}P_{\rm GF}^{1.43}$ (equations ~\ref{eq:M_ej,max,loc},\ref{eq:M_ej,min,loc}; Fig.~\ref{fig:M_ej}), if one instead fixes the flare energy $E_{\rm GF} \propto f_{\rm open} P_{\rm GF}$ (equation \ref{eq:E_GF}), then $M_{\rm ej} \propto E_{\rm GF}^{1.43}f_{\rm open}^{-0.43}$.  Thus, confining a flare of the same total energy to a smaller and smaller covering fraction of the NS surface (smaller $f_{\rm open} \ll 1$), increases the total ejecta mass, potentially bringing the ejecta mass implied by the radio afterglow of the Dec. 2004 GF from SGR 1806-20 into better accord with the energy budget implied by the radiated gamma-rays (Sec.~\ref{sec:afterglow}).  

How realistic is the sudden appearance of a high-pressure region above a localized region of the NS surface?  At least the ``sudden'' aspect of such a picture is supported by the rapid rise-time of the gamma-ray light curves $\lesssim 0.25$ ms (e.g., \citealt{palmer2005}), comparable to the light crossing time of the magnetosphere (e.g., \citealt{lyutikov2006}).  The source of dissipated energy may be Alfv\'{e}n waves launched by a substantial crustal failure \citep[e.g.,][]{perna2011,lander2015} or an otherwise source of large-scale reconnection of magnetic field \citep[e.g.,][]{parfrey2013,lander2016}.  

While a 1D calculation may be a reasonable approximation to the physical picture if $f_{\rm open}$ is not too small, confining the dissipated energy source to a narrower and narrower patch of the NS surface (in essence, envisioning a highly-localized ``magnetic bomb'') must eventually violate a 1D picture.  Indeed, a natural expansion of our approach would be to increase the number of spatial dimensions of our numerical simulations to two or three.  The inclusion of additional physical effects$-$magnetic fields, neutrino cooling, radiation transport, and general relativity$-$would also add more realism to the set-up.  Predictions for the ejecta nucleosynthesis could also be improved using nuclear reaction network calculations.  We leave these important extensions to future work.

\section*{Acknowledgements}

JC thanks Ond\v{r}ej Pejcha for useful comments.  JC is supported by Horizon 2020 ERC Starting Grant "Cat-In-hAT" (grant agreement no. 803158) and by the Charles University Mobility Fund (proposal no. FM/c/2022-2-048).  TAT is supported in part by NASA grant 80NSSC20K0531. BDM is supported in part by the National Science Foundation (grant \# AST-2002577).  The Flatiron Institute is supported by the Simons Foundation.

Most of the algebraic calculations and visualizations in this work were performed with \textsc{Mathematica 12.2} \citep{mathematica12.2}, \textsc{Matplotlib} \citep{hunter2007}, \textsc{NumPy} \citep{harris2020}.

\section*{Data Availability}

The data underlying this article will be shared on reasonable request to the corresponding author.

The videos produced from the Fiducial model are publicly available in \textsc{Zenodo}, at \href{https://dx.doi.org/10.5281/zenodo.10593900}{https://dx.doi.org/10.5281/zenodo.10593900}, on \textsc{YouTube}, at \href{https://www.youtube.com/playlist?list=PLtfS1xXqUmtC2A48QhRvGXxbeMK1vkNnC}{youtube.com/playlist?list=PLtfS1xXqUmtC2A48QhRvGXxbeMK1vkNnC}, and alongside the paper as supplementary material on the journal website.



\bibliographystyle{mnras}
\bibliography{bibliography} 




\appendix

\onecolumn

\section{Analytic estimates for mass-weighted ejecta distributions}\label{app:analytic}

\begin{figure*}
	\includegraphics[width=\textwidth]{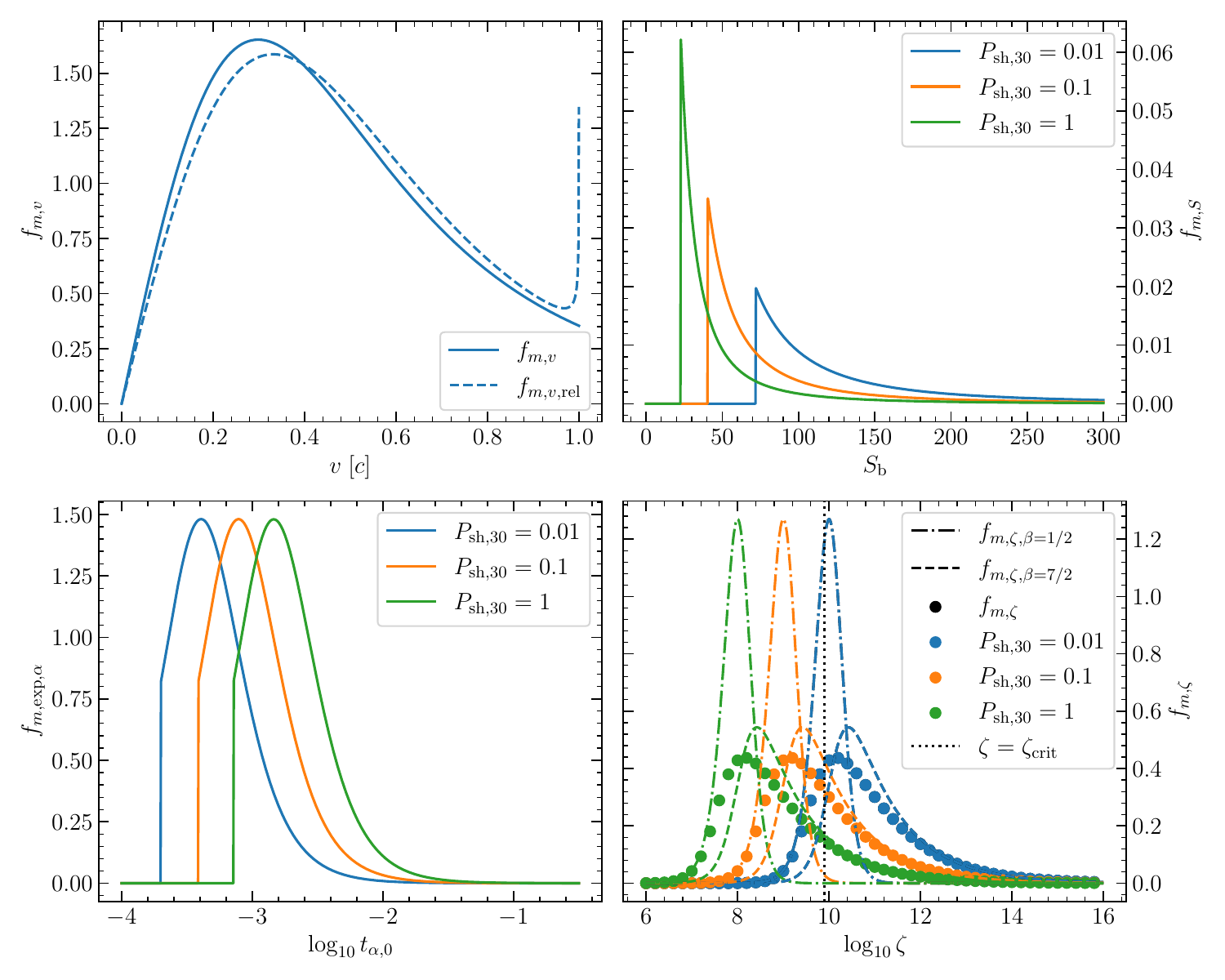}
    \caption{Analytically derived mass distributions for different unbound ejecta properties for three different values of the pressure of the shock-heated material $P_{\rm sh}$ as labeled.  We show the classical $f_{m,v}$ (equation \ref{eq:f_m,v}) and the relativistic $f_{m,v,\rm{rel}}$ (equation \ref{eq:f_m,v,rel}) velocity distributions, the distribution of the expansion timescale $t_\alpha$ at the $\alpha$-formation radius $f_{m,\rm{exp},\alpha}$ (equation \ref{eq:f_m,exp,alpha}), and the entropy distribution $f_{m,S}$ (equation \ref{eq:f_m,S}).  We also show the numerical $\zeta$-distribution $f_{m,\zeta}$ together with an approximate analytic solution $f_{m,\zeta,\beta}$ (equation \ref{eq:f_m,beta,zeta,alpha}).  The distributions are normalized so that $\int_{- \infty}^{\infty} f_{m,i}(x) \dd x = 1$.}
    \label{fig:hist_an}
\end{figure*}

As a first crude approximation we can assume that the total energy of a shock-heated crustal shell is conserved for each shell separately.  At the onset of the GF this energy is composed of only the internal and gravitational potential components, while at the end of the simulation only the kinetic component remains.  Hence, equation (\ref{eq:en_dens}) implies
\begin{equation}\label{eq:en_class}
    \frac{1}{\Gamma-1} \frac{P_{\rm sh}}{\rho_{\rm sh}} - G \frac{M_\ns}{R_\ns} = \frac{1}{2} v^2,
\end{equation}
where again $\rho_{\rm sh} = 7 \rho_{\rm cr}$ and $P_{\rm sh}$ is given by equation (\ref{eq:P_sh}).  Equations (\ref{eq:polytrope},\ref{eq:M_cr}) define the crustal density as function of the crustal mass coordinate:
\begin{equation}\label{eq:rho_m}
    \rho_{\rm cr} = \lp \frac{G M_\ns}{4 \pi R_\ns^4} \frac{M_{\rm cr}}{P_*}\rp^{\frac{1}{\Gamma_*}} \rho_*.
\end{equation}
The cumulative velocity-dependent mass distribution that gives the unbound ejecta mass with velocity $v^\prime \in \langle 0, v\rangle$ is defined by
\begin{equation}
    F_{m,v} (v) = M_{\rm ej,loc} - M_{\rm cr} (v), \quad \text{for} \quad v \geq 0,
\end{equation}
where we assume that the total ejecta mass $M_{\rm ej}$ is given by the \emph{local} energy criterion (equation \ref{eq:M_ej,max,loc}) if we consider that the pressure of the shock-heated material $P_{\rm sh}$ could in general be lower than $P_{\rm GF}$ (equation \ref{eq:P_sh}), i.e.
\begin{equation}\label{eq:M_ej,loc}
    M_{\rm{ej,loc}} \equiv M_{\rm ej,max,loc} (P_{\rm GF} \rightarrow P_{\rm sh}) = \frac{4 \pi R_\ns^4}{G M_\ns} P_* \lp  \frac{1}{7} \frac{P_{\rm{sh}}/(\Gamma-1)}{G M_\ns / R_\ns} \frac{1}{\rho_*}\rp^{\Gamma_*} \approx 8.5 \times 10^{25} \: \text{g} \:  P_{\rm{sh},30}^{1.43},
\end{equation}
and $M_{\rm cr}(v)$ is given by equations (\ref{eq:en_class},\ref{eq:rho_m}).  It holds $F_{\text{m},v} (0) = 0$ and (since this treatment neglects special relativistic effects) $\lim_{v \to \infty} F_{\text{m},v} (v) = M_{\rm ej,loc} = M_{\rm ej}$.  The velocity-dependent mass distribution of the ejecta is then given by
\begin{equation}\label{eq:f_m,v}
    f_{m,v}(\tilde{v}) \propto \frac{\dd F_{m,v}}{\dd v} \propto \frac{\tilde{v}}{\lp \tilde{v}^2 + \tilde{v}_{\rm esc}^2\rp^{\Gamma_* + 1}}, \quad \rm{for} \quad \tilde{v} \geq 0,
\end{equation}
where $\tilde{v} = v / c$ and
\begin{equation}\label{eq:tilde_v_esc}
    \tilde{v}_{\rm esc}^2 = \frac{R_{\rm SCH}}{R_\ns} = \frac{2 G M_\ns}{c^2 R_\ns}.
\end{equation}
The distribution peaks at
\begin{equation}
    \left. \frac{\dd f_{m,v}}{\dd \tilde{v}} \right\vert_{\tilde{v} = \tilde{v}_{\rm peak}} = 0, \quad \rm{and} \quad \tilde{v}_{\rm peak} = \frac{\tilde{v}_{\rm esc}}{\sqrt{2 \Gamma_* + 1}} = \frac{0.587}{1.965} = 0.299.
\end{equation}
We can generalize the distribution to the special-relativistic case by changing the kinetic term in equation (\ref{eq:en_class}) as follows
\begin{equation}\label{eq:en_rel}
    \frac{1}{\Gamma-1} \frac{P_{\rm sh}}{\rho_{\rm sh}} - G \frac{M_\ns}{R_\ns} = (\gamma-1) c^2, 
\end{equation}
where $\gamma = (1-v^{2}/c^{2})^{-1/2}.$  In this case the relativistic velocity-dependent mass distribution reads
\begin{equation}\label{eq:f_m,v,rel}
    f_{m,v,\rm{rel}} (\tilde{v}) \propto \frac{\tilde{v} \lp 1- \tilde{v}^2\rp^{\frac{\Gamma_*}{2}-1}}{\left[ 1- \lp 1 - \frac{1}{2} \tilde{v}_{\rm esc}^2\rp \sqrt{1 - \tilde{v}^2}\right]^{\Gamma_* + 1}}, \quad \rm{for} \quad \tilde{v} \in \langle 0,1\rangle.
\end{equation}
The two distributions, $f_{m,v}$ and $f_{m,v,\rm{rel}}$, are very similar with the exception that $f_{m,v,\rm{rel}}$ has a singularity at $\tilde{v} = 1$, see Fig.~\ref{fig:hist_an}.  Thus, we can use the non-relativistic distribution $f_{m,v}$ for $\tilde{v} \in \langle 0,1\rangle$.

Next, assuming that at late times each shell reaches a constant terminal velocity, and taking into account the definition of the expansion timescale (\ref{eq:t_exp}) we immediately see that the mass distribution of $t_{\rm exp}$ is a $\delta$ distribution centered at the simulation time
\begin{equation}
    f_{m,\rm{exp}} (t_{\rm exp}) = \delta (t_{\rm exp} - t_{\rm sim}).
\end{equation}
Using this knowledge we can calculate the expansion timescale corresponding to the $\alpha$-formation radius $R_\alpha$ (equation \ref{eq:R_alpha}) as
\begin{equation}\label{eq:t_exp,alpha_app}
    t_{\alpha,0} = \frac{R_{\alpha,0} - R_{\ns,0}}{v_0},
\end{equation}
where $R_0 = R / (1 \: \rm{cm})$ and $v_0 = v / (1 \: \rm{cm} \: \rm{s}^{-1})$.  Assuming constant entropy ($S_{\rm b} \propto T^{3}/\rho$; equation \ref{eq:S_b}) and homologous expansion ($\rho(r) \propto r^{-3}$) yields $T(r) \propto r^{-1}$.  Because the temperature of the shocked-heated material obeys,
\begin{equation}\label{eq:T_shocked}
    T_{\rm sh} \simeq \lp \frac{12 P_{\rm sh}}{11 a}\rp^{1/4},
\end{equation}
the $\alpha$-formation radius becomes
\begin{equation}\label{eq:R_alpha_app}
    R_\alpha = \frac{T_{\rm{sh,MeV}}}{T_{\alpha,\rm{MeV}}} R_\ns = 2T_{\rm{sh,MeV}} R_\ns \approx 18.9 R_\ns P_{\rm{sh},30}^{1/4}.
\end{equation}
From equations (\ref{eq:f_m,v},\ref{eq:t_exp,alpha_app}) we can express the mass distribution of the expansion timescale at the $\alpha$-formation radius as
\begin{equation}\label{eq:f_m,exp,alpha}
    f_{m,\rm{exp},\alpha} (\log_{10} t_{\alpha,0}) \propto \frac{10^{-2 \log_{10} t_{\alpha,0}}}{\left[ \lp \frac{R_{\alpha,0} - R_{\ns,0}}{c_0}\rp^2 10^{-2 \log_{10} t_{\alpha,0}} + \tilde{v}_{\rm esc}^2\right]^{\Gamma_* + 1}}, \quad \rm{for} \quad t_{\alpha,0} \geq \frac{R_{\alpha,0} - R_{\ns,0}}{c_0},
\end{equation}
where $R_\alpha$ is given by equation (\ref{eq:R_alpha_app}) and $\tilde{v}_{\rm esc}$ by equation (\ref{eq:tilde_v_esc}).

Further, assuming that shocked crustal material is heated to a constant temperature $T_{\rm sh}$ (equation \ref{eq:T_shocked}), its entropy is given by (equation \ref{eq:S_b}) 
\begin{equation}\label{eq:S_b_shocked}
    S_{\rm b} = 5.21 T_{\rm sh, MeV}^3 \rho_{\rm{sh,8}}^{-1}.
\end{equation}
The minimum entropy of the unbound ejecta $S_{\rm b,min}$ corresponds to those layers with the maximal pre-shock density $\rho_{\rm cr,max}$ which just become marginally unbound.  Obtaining the latter from  equation (\ref{eq:en_class}) for $v = 0$, this yields
\begin{equation}
    S_{\rm b,min} = 22.7 P_{\rm{sh},30}^{-1/4}.
\end{equation}
Equations (\ref{eq:rho_m}, \ref{eq:S_b_shocked}) define the entropy-dependent mass distribution:
\begin{equation}\label{eq:f_m,S}
    f_{m,S} (S_{\rm b}) \propto \frac{1}{S_{\rm b}^{\Gamma_* + 1}}, \quad \rm{for} \quad S_{\rm{b}} \geq S_{\rm b,min}.
\end{equation}

Combining equations (\ref{eq:en_class},\ref{eq:rho_m},\ref{eq:M_ej,loc},\ref{eq:t_exp,alpha_app},\ref{eq:T_shocked},\ref{eq:R_alpha_app},\ref{eq:S_b_shocked}) we can express the $r$-process figure-of-merit parameter $\zeta$ (equation \ref{eq:zeta}) as
\begin{equation}\label{eq:zeta_sigma}
    \zeta (\sigma) = Z \frac{\lp 1-\sigma \rp^{1/2}}{\sigma^{7/2}},
\end{equation}
where
\begin{equation}\label{eq:Z_sigma}
    Z = 2.15 \times 10^9 \: Y_{\rm{e},43}^{-3} \frac{P_{\rm{sh},30}^{-3/4}}{18.9 P_{\rm{sh},30}^{1/4}-1} \approx 1.2 \times 10^8 \: Y_{\rm{e},43}^{-3} P_{\rm{sh},30}^{-1}, \quad \text{and} \quad \sigma = \lp \frac{M_{\rm cr}}{M_{\rm ej,loc}} \rp^{\frac{1}{\Gamma_*}},
\end{equation} $Y_{\rm{e},43} = Y_{\rm e}/0.43$ and we have assumed $Y_{\rm e}$ is constant in the unbound ejecta.  It also holds $\sigma \in \langle 0,1 \rangle$.  In order to calculate the $\zeta$-dependent mass distribution we need to invert the function $\zeta (\sigma)$.  This leads to a septic equation: $(\zeta^2 / Z^2) \sigma^7+ \sigma - 1 = 0$.  We are not aware of the existence of an algebraic solution to this equation.  Thus, to proceed analytically we use an approximation
\begin{equation}
    \zeta (\sigma) \approx Z \frac{\lp 1-\sigma^{2 \beta} \rp^{\frac{1}{2}}}{\sigma^{\beta}},
\end{equation}
where $\beta \in \langle \frac{1}{2}, \frac{7}{2}\rangle$.  If $\beta = 1/2$ then the numerator is the same and we capture the limiting behaviour $\sigma \rightarrow 1$ correctly, if $\beta = 7 / 2$ then the denominator is the same and we capture $\sigma \rightarrow 0$ correctly.  The inverse function in this approximation reads
\begin{equation}
    \sigma (\zeta) = \frac{1}{\lp 1 + \zeta^2 / Z^2\rp^{\frac{1}{2 \beta}}}.
\end{equation}
Using equation (\ref{eq:Z_sigma}) then leads to the $\zeta$-dependent mass distribution in the form
\begin{equation}\label{eq:f_m,beta,zeta,alpha}
    f_{m,\zeta} (\log_{10} \zeta) \sim f_{m,\zeta,\beta} (\log_{10} \zeta) \propto \frac{10^{2 \log_{10} \zeta}}{\lp Z^2 + 10^{2 \log_{10} \zeta} \rp^{\frac{\Gamma_*}{2 \beta} + 1}}.
\end{equation}
We also solve equation (\ref{eq:zeta_sigma}) numerically and are able to compute the $\zeta$-dependent mass distribution $f_{m,\zeta} (\log_{10} \zeta)$ stemming from this equation precisely.

Fig.~\ref{fig:hist_an} shows the distributions from equations (\ref{eq:f_m,v},\ref{eq:f_m,v,rel},\ref{eq:f_m,exp,alpha},\ref{eq:f_m,S},\ref{eq:f_m,beta,zeta,alpha}).  We see that $f_{m,\zeta,\beta = 1/2}$ approaches the numerical solution $f_{m,\zeta}$ for $\zeta \ll Z$, i.e. $\sigma \rightarrow 1$ (equation \ref{eq:zeta_sigma}), and $f_{m,\zeta,\beta = 7/2}$ approaches $f_{m,\zeta}$ for $\zeta \gg Z$, i.e. $\sigma \rightarrow 0$, as expected.  A comparison of Fig.~\ref{fig:16.0-1.0-ls-hist} and Fig.~\ref{fig:hist_an} reveals the mass-weighted distributions exhibit qualitatively similar behaviours.  The total ejecta mass in the Fiducial case (Fig.~\ref{fig:16.0-1.0-ls-hist}) is $M_{\rm ej} \approx 10^{25.5} \: \rm{g}$ (Table~\ref{tab:results}), which according to equation (\ref{eq:M_ej,loc}) corresponds to $P_{\rm{sh},30} \simeq 0.5$ or equivalently $\alpha \simeq 1/8$.  Hence, the distributions in Fig.~\ref{fig:16.0-1.0-ls-hist} should be compared to the green (possibly orange) lines (dots) in Fig.~\ref{fig:hist_an}.  We observe quantitative differences between these two sets of distributions; namely, the analytical results give greater $v$ approaching $c$, longer $t_{\alpha,0}$ with a greater spread, lower $S_{\rm b}$, and lower $\zeta$.  We suppose these differences between the analytic estimates and numerical simulations to arise from further idealisations made in order to arrive at an analytic solution.  These idealisations include: (1) neglecting the rarefaction wave originating from the outer edge of the high-pressure shell ($M_{\rm ej,min,loc}$; equation \ref{eq:M_ej,min,loc}), hence effectively assuming an infinitely long high-pressure shell; (2) not allowing the escaping layers to exchange energy; and (3) assuming constant entropy and homologous expansion all the way from $r = R_\ns$.  Indeed, we can consider the analytic case with $\alpha \approx 1$, i.e. $P_{\rm sh,30} \approx P_{\rm GF,30} \approx 4$.  This yields $M_{\rm ej,loc} = 6.1 \times 10^{26} \: \rm{g}$ (equation \ref{eq:M_ej,loc}), but we can account for the rarefaction wave by assuming that only the most energetic layers with mass of $M_{\rm ej} \approx 10^{25.5} \: \rm{g}$ get ejected.  Fig.~\ref{fig:S_b} implies that these layers have entropy of $S_{\rm b} \gtrsim 10^2$, in line with our numerical results (Fig.~\ref{fig:16.0-1.0-ls-hist}).  In this scenario, the entropy distribution is ``corrected" but the velocity distribution remains ``skewed" since the most energetic layers have velocity approaching $c$.  To ``correct" the velocity distribution as well would require allowing energy to be exchanged between different layers, but it is not clear how to account for this analytically.  As none of the above idealisations apply to the numerical results, we believe them to be more reliable than the the analytic estimates.  Nevertheless, the analytic results yield the correct qualitative behaviour and still provide valuable insight.

\section{Distributions of unbound ejecta properties for different simulation setups}\label{app:hist_comp}

\begin{figure*}
	\includegraphics[width = 0.98\textwidth]{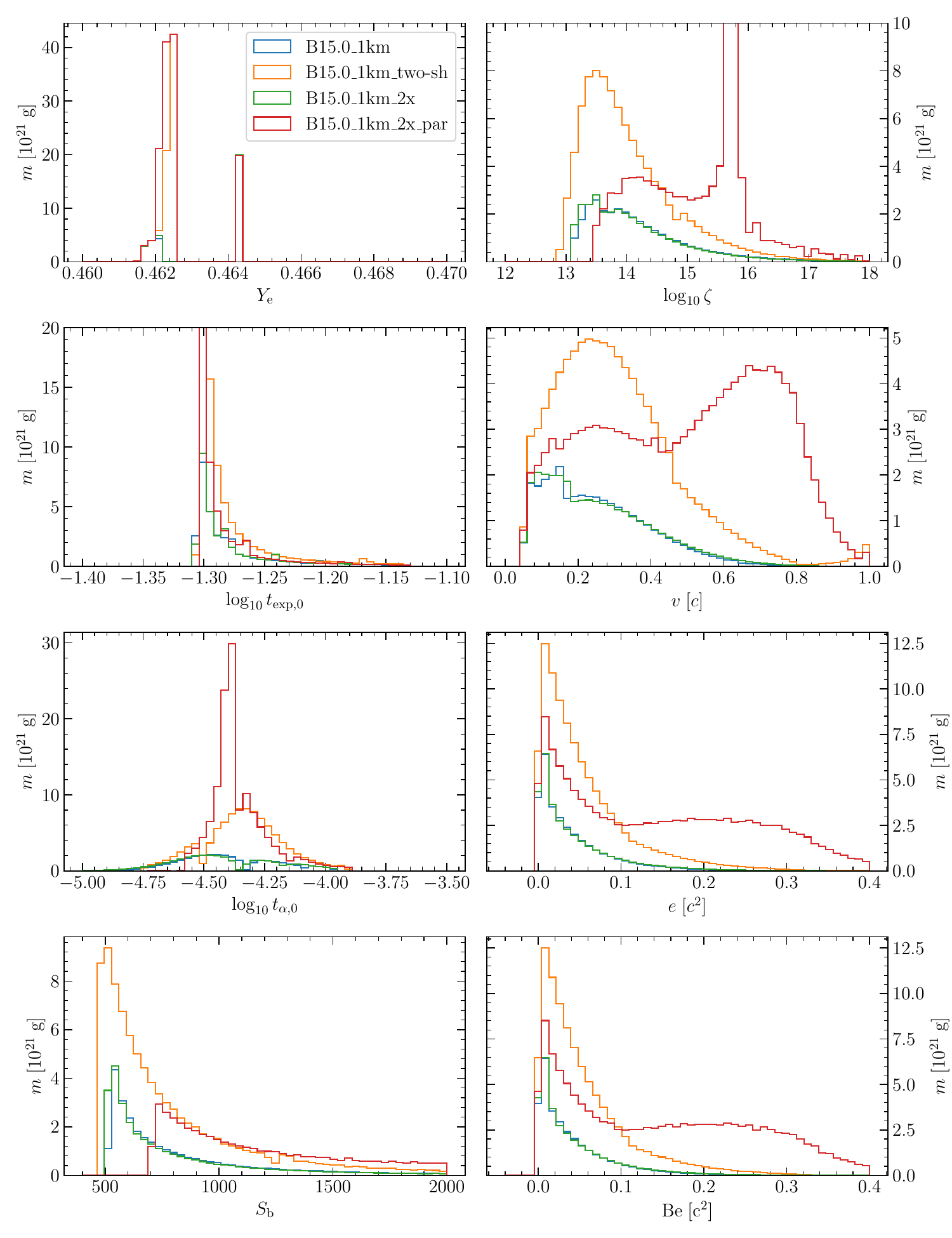}
    \caption{Mass-weighted distributions of properties of the unbound ejecta mass (i.e. only those layers satisfying $e > 0$), shown at $t = 50.0 \: \rm{ms}$ for four different models as labeled.  The quantities shown are the same as in Fig.~\ref{fig:16.0-1.0-ls-hist}.}
    \label{fig:hist_comp}
\end{figure*}

In Fig.~\ref{fig:hist_comp}, we show the mass-weighted distributions of key quantities for four different convergence test models.  The two models against which we test are B15.0\_1km and B15.0\_1km\_2x.  The two models that show the greatest differences and thus we show them are B15.0\_1km\_two-sh and B15.0\_1km\_2x\_par, see Table~\ref{tab:convergence}.  We see that B15.0\_1km and B15.0\_1km\_2x show almost identical results.  Both, B15.0\_1km\_two-sh and B15.0\_1km\_2x\_par, diverge considerably, but B15.0\_1km\_two-sh keeps the same qualitative behaviour, while B15.0\_1km\_2x\_par differs qualitatively from the remaining three models.  We see for example the double-peaked velocity distribution for which we do not see any physical reason.  Thus, we disfavour B15.0\_1km\_2x\_par.  Based on the differences between B15.0\_1km and B15.0\_1km\_two-sh we can conclude that $M_{\rm ej}$ is constrained to around half an order of magnitude by the numerical simulations.


\bsp	
\label{lastpage}
\end{document}